\documentclass[twocolumn,pre,superscriptaddress]{revtex4}
\usepackage{tikz}
\usepackage{graphicx}%
\usepackage{color} 
\usepackage{transparent}
\usepackage{psfrag}
\usepackage{dcolumn}
\usepackage{amsmath}
\usepackage{amsmath,bm} 
\usepackage{amssymb}
\usepackage{latexsym}
\usepackage{hyperref} 
\usepackage{booktabs}
\usepackage{latexsym, bbold}
\begin{document}

\newcommand{\tikzcircle}[2][red,fill=red]{\tikz[baseline=-0.5ex]\draw[#1,radius=#2] (0,0) circle ;}%
\def\bea{\begin{eqnarray}}
\def\eea{\end{eqnarray}}
\def\beq{\begin{equation}}
\def\eeq{\end{equation}}
\def\f{\frac}
\def\k{\kappa}
\def\e{\epsilon}
\def\ve{\varepsilon}
\def\be{\beta}
\def\D{\Delta}
\def\h{\theta}
\def\t{\tau}
\def\a{\alpha}

\def\cDa{{\cal D}[X]}
\def\cD{{\cal D}[x]}
\def\cL{{\cal L}}
\def\cLo{{\cal L}_0}
\def\cLa{{\cal L}_1}

\def\Re{{\rm Re}}
\def\sj{\sum_{j=1}^2}
\def\rk{\rho^{ (k) }}
\def\rek{\rho^{ (1) }}
\def\cek{C^{ (1) }}
\def\rz{\rho^{ (0) }}
\def\rt{\rho^{ (2) }}
\def\rtb{\bar \rho^{ (2) }}
\def\trk{\tilde\rho^{ (k) }}
\def\trek{\tilde\rho^{ (1) }}
\def\trz{\tilde\rho^{ (0) }}
\def\trt{\tilde\rho^{ (2) }}
\def\r{\rho}
\def\tD{\tilde {D}}

\def\rpl{r_\parallel}
\def\rp{{\bf r}_\perp}

\def\s{\sigma}
\def\kb{k_B}
\def\bF{\bar{\cal F}}
\def\F{{\cal F}}
\def\la{\langle}
\def\ra{\rangle}
\def\nn{\nonumber}
\def\up{\uparrow}
\def\dn{\downarrow}
\def\S{\Sigma}
\def\dg{\dagger}
\def\d{\delta}
\def\p{\partial}
\def\l{\lambda}
\def\L{\Lambda}
\def\G{\Gamma}
\def\o{\Omega}
\def\w{\omega}
\def\g{\gamma}

\def\bv{ {\bf b}}
\def\uv{ {\hat{\bm{u}}}}
\def\rv{ {\bf r}}
\def\vv{ {\bf v}}

\def\jv{ {\bf j}}
\def\jr{ {\bf j}_r}
\def\jd{ {\bf j}_d}
\def\jdd{ { j}_d}
\def\noi{\noindent}
\def\a{\alpha}
\def\d{\delta}
\def\p{\partial} 

\def\la{\langle}
\def\ra{\rangle}
\def\e{\epsilon}
\def\n{\eta}
\def\g{\gamma}
\def\break#1{\pagebreak \vspace*{#1}}
\def\hf{\frac{1}{2}}

\title{Active Brownian particles:  mapping to equilibrium polymers and  exact computation of moments }
\author{Amir Shee}
\email{amir@iopb.res.in}
\affiliation{Institute of Physics, Sachivalaya Marg, Bhubaneswar 751005, India}
\affiliation{Homi Bhaba National Institute, Anushaktigar, Mumbai 400094, India}

\author{Abhishek Dhar}
\email{abhishek.dhar@icts.res.in}
\affiliation{International Centre for Theoretical Sciences, Tata Institute of Fundamental Research, Bengaluru 560089, India}

\author{Debasish Chaudhuri}
\email{debc@iopb.res.in}
\affiliation{Institute of Physics, Sachivalaya Marg, Bhubaneswar 751005, India}
\affiliation{Homi Bhaba National Institute, Anushaktigar, Mumbai 400094, India}

\date{\today}

\begin{abstract}
It is well known that path probabilities of Brownian motion correspond to the equilibrium configurational probabilities of flexible Gaussian polymers, while those of active Brownian motion correspond to in-extensible semiflexible polymers.  
Here we investigate the properties of the equilibrium polymer that  corresponds to the trajectories of  particles acted on simultaneously by both Brownian as well as  active noise. 
Through this mapping we can see interesting crossovers in mechanical properties of the polymer with changing contour length. The polymer end-to-end distribution exhibits  Gaussian behaviour for short lengths, which  changes to the form of semiflexible filaments at  intermediate lengths, to finally go back to a Gaussian form for long contour lengths. By performing a Laplace transform of the governing Fokker-Planck  equation of the active Brownian particle, we discuss a direct method to derive exact expressions for all the moments of the relevant dynamical variables, in arbitrary dimensions. These are verified via numerical simulations and used to describe interesting qualitative features such as, for example,  dynamical crossovers. 
Finally we discuss the kurtosis of the ABP's position which we compute exactly and show that it can be used to differentiate between active Brownian particles and  active Ornstein-Uhlenbeck process.   
\end{abstract}

\maketitle

\section{Introduction}

Active particles are  entities that can perform dissipative self-propulsion even in the absence of external driving force.  
Their dynamics violates equilibrium fluctuation-dissipation relation. The energy required for the motion 
is supplied at the local scale by different processes depending on the situation, e.g., internal energy depot in bacteria, hydrolysis of chemical fuel like ATP for molecular motors, and 
transverse shaking in active granular matter~\cite{Schweitzer2003, Bechinger2016, Marchetti2013, Romanczuk2012, Vicsek2012, Julicher1997b}.  The direction of active motion is decided
by the inbuilt asymmetry of the particles~\cite{Vicsek2012}, or environment, e.g., as provided by filamentous tracks for molecular motors~\cite{Julicher1997b}. 
The system remains out of equilibrium, detailed balance being broken naturally in self propulsion.

 The model of active Brownian particles (ABP), in which a particle has its own heading direction of self-propulsion, while the heading direction itself performs rotational diffusion, has been used to describe self-propelled colloidal particles \cite{howse2007,palacci2010}. Its behavior becomes equivalent to that of bacterial run and tumble motion in the long time limit~\cite{Cates2012}. A related model of the active Ornstein-Uhlenbeck process (AOUP) also describes self propulsion and has attracted considerable attention recently, due to its relative simplicity ~\cite{Fodor2016, Kurzthaler2018, Das2018}.

Despite a tremendous advancement in the knowledge of collective properties of active matter, the dynamics of single active particles  is not yet 
completely understood. Some recent  analytic results~\cite{sevilla2014,sevilla2015, Kurzthaler2016,Kurzthaler2018,Wagner2017,Pototsky2012, Duzgun2018, Basu2018, Basu2019,Das2018,Malakar2018,maes2018,Malakar2020,Dhar2019} indicate 
the qualitatively rich physics that even a single active particle can exhibit.
The work in \cite{sevilla2014,sevilla2015} considered free ABP in two dimensions in the presence of thermal noise. Using a Fourier series expansion of the corresponding Fokker-Planck equation they were able to obtain various analytic results for the radial distribution and also some moments. In particular they computed the Kurtosis and pointed out that this could be used to differentiate the ABP from Gaussian models such as the AOUP. The same model was solved exactly in \cite{Kurzthaler2018} by using a series expansion involving a Fourier basis for the position of the ABP and a Mathieu functions basis for its angular degree of freedom. On the other hand \cite{Basu2018,Basu2019} studied  ABP without thermal noise and obtained exact results for short time and long time asympotitic properties of the positional distributions and pointed out the presence of anisotropies in short time behaviour.

Remarkably, the Fokker-Planck equation corresponding to the ABP, in the absence of thermal noise, was studied as early as 1952~\cite{Hermans1952,Daniels1952} in the context of understanding the so-called worm-like-chain (WLC) model of semi-flexible polymers. The WLC model is the continuous version of the Kratky-Porod model, which in turn corresponds to a persistent random walker. In fact an exact mapping can be obtained between the trajectories of an ABP and the equilibrium configurations of the semi-flexible polymer and this has been used to understand the equilibrium properties of the polymer~\cite{Dhar2002,Chaudhuri2007,CastroVillarreal2019}. On the other hand it is well known that trajectories of passive Brownian particles generate so-called flexible Gaussian  polymers~\cite{Doi1986}. 
In terms of energetics, the WLC model is one which has only bending energy, while the Gaussian polymer  has  only stretching energy. It is then natural to ask what the  polymer model  would be that corresponds to an ABP in the presence of translational thermal noise. One of the aims of the paper is to explore this connection. A second main objective of the paper is to point out that the approach of \cite{Hermans1952} provides an efficient method of computing all moments (of both positional and orientational degrees of freedom)  of the ABP (with or without translational thermal noise), in arbitrary dimensions.

In this paper, we consider free ABPs in $d$-dimensions, in the presence of translational thermal noise. We summarize here our main results:-

1) We discuss the mapping of the ABP trajectories to the equilibrium polymer configurations.   We point out that the polymer model differs from the physical system including both bending and stretching energy and the physical relevance is thus not clear.  
Nevertheless we illustrate the mapping by comparing  results for various dynamical moments and displacement distribution functions of ABPs with  the corresponding  polymer properties obtained from equilibrium polymer simulations. 
We show that ABP simulations provide an efficient alternate means of obtaining equilibrium polymer properties that usually require Monte-Carlo or Langevin simulations. 

2) We  show how arbitrary  moments of position and orientation degrees of freedom can be computed exactly by utilizing the Fokker-Planck equation governing the dynamics of ABP, using the approach in \cite{Hermans1952}. Interesting dynamical crossovers displayed by the moments are analyzed using the exact expressions.  The resultant dynamics crosses over from short-time equilibrium diffusion, to intermediate time  active ballistic motion, to long time effective diffusion. Short time anisotropies in the distribution, pointed out in \cite{Basu2018} are also discussed.  As has been pointed out in earlier studies \cite{sevilla2014,Das2018}, we show that the Kurtosis, which we compute exactly, can be used to distinguish  between the ABP and the AOUP models. 

The plan of the paper is as follows.  In Sec.\,\ref{sec:model}, we present the ABP model in the presence of translational diffusion. We demonstrate the mapping of ABP trajectories to an equilibrium polymer in Sec.\,\ref{sec:poly}. 
In this section, using polymer simulations we show the comparison of the results between the two models. In Sec.\,\ref{sec:moments} we present the detailed analytical calculation of the moments, and use them to analyze the simulation results. In Sec.~\ref{sec:chainL} we present a detailed analysis of how the distribution function of displacement of ABP changes with time. Equivalently, this shows how the behavior of the related polymer model changes from a Gaussian to a semiflexible filament with increase in contour length of the chain. Finally, in Sec.~\ref{sec:aoup}, the properties of ABP and AOUP are distinguished. We provide an exact expression of the generalized Kurtosis, which can be utilized in analyzing experimental results of self propelled colloids to decide whether AOUP would suffice or one needs to invoke ABP to describe the observed behavior.  We conclude in Sec.\,\ref{sec:outlook} with a discussion of our main results.    

 
 \begin{figure}[!t]
\begin{center}
\includegraphics[width=8.6cm]{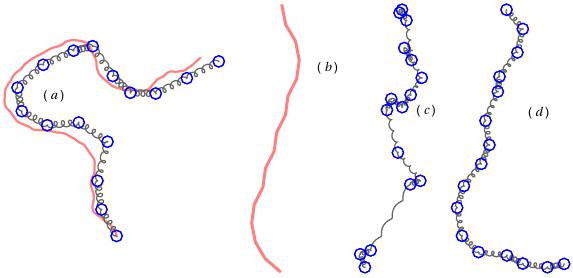} 
\caption{(color online) Typical configurations of 
($a$)~the polymer mapping of the ABP model (Eq.\eqref{eq:H}),
($b$)~the related worm-like-chain (Eq.\eqref{eq:wlc}),
($c$)~the related Gaussian polymer under a directed external force (Eq.\eqref{eq:Gauss}), and
($d$)~the related extensible semiflexible chain (Eq.\eqref{eq:esc}). 
The thick (red) line in ($a$) shows the worm-like-chain conformation that the Gaussian polymer denoted by the beads and springs tries to align with. 
The configurations are plotted with chain length $L=15\,\s$, persistence length $\ell_p = 4.28\,\s$, spring constant $A=30.0\, \s^{-1}$ in ($a$) and ($d$), and $A=1.0\, \s^{-1}$ in ($c$) . 
Note that $A$  does not play any role in deciding the conformation in ($b$). 
} 
\label{fig:conf}
\end{center}
\end{figure}

\section{Definition of model}
\label{sec:model}
{
 The active Brownian particle (ABP) in $d$-dimensions is described by its position $\rv = (r_1,r_2,\ldots,r_d)$ and its orientation $\uv=(u_1,u_2,\ldots,u_d)$ which is a unit vector in $d$-dimensions. Let the infinitesimal increments at time $t$ are denoted by $d r_i  = r_(t+dt) - r_i(t)$ and $d u_i  = u_i(t+dt) - u_i(t)$. In Ito convention, the equation of motion of the ABP is given by
 \bea
d r_i &=& v_0 u_i \, dt + {dB_i^{\rm t}}(t) \label{eq:eom1}
\\
d u_i &=&  (\delta_{ij} -u_i u_j)\, dB_j^{\rm r}-{(d-1){D}_r} u_i dt,  \label{eq:eom2}
 \eea
where the Gaussian noise terms ${\bm {dB}}^{\rm t}$ and ${\bm {dB}}^{\rm r}$  have mean zero and variances  $\la {dB_i^{\rm t}}  {dB_j^{\rm t}} \ra = 2 D \delta_{ij} dt$, $\la {dB_i^{\rm r}}  {dB_j^{\rm r}} \ra =  2 {D}_r \d_{ij} dt$ control the translational and rotational diffusions, respectively. Alternatively, we can write Eq.\ref{eq:eom2} in the Stratonovich form 
\bea
(S)~~ d u_i =  (\delta_{ij} -u_i u_j) \circ \eta_j^{\rm r} \, dt. \nn
\label{eq_S}
\eea 
Eq.(\ref{eq:eom1}) gives $\la d r_i \ra = v_0 u_i dt$ and $\la d r_i d r_j \ra = 2 D \d_{ij} dt$.} 
The form of Eq.\ref{eq:eom2} ensures the normalization $\uv^2=1$ at all times. Eq.\ref{eq:eom2} implies the mean and variance of orientational fluctuations 
\bea
\la d u_i \ra = -(d-1) D_r \, u_i\, dt
\eea
 and 
 \bea
 \la d u_i du_j \ra = 2D_r (\d_{ij} - u_i u_j)\, dt. 
 \label{eq:rot}
 \eea
 These results are utilized in deriving the  Fokker-Planck  equation for this system which is discussed in Sec.\ref{sec:moments}.

It is straightforward to perform a direct numerical simulation of Eq.s(\ref{eq:eom1}), (\ref{eq:eom2}) using the Euler-Maruyama integration scheme to generate trajectories of motion. In the following,  we first show how the ABP trajectories can be mapped to an effective polymer model.

\section{Mapping to equilibrium polymer}
\label{sec:poly}
The probability distribution of a stochastic trajectory $\{\rv(t),\uv(t)\}$, corresponding to the evolution Eq.(\ref{eq:eom1}) and (\ref{eq:eom2}) over the time-range $t \in (0,\t)$, is given by~\cite{hsustochastic, kleinert2009path} 
\bea
{\cal P} [ \{\rv (t), \uv (t)\}] \propto e^{- \f{1}{4D} \int_0^\t dt \left( \f{\p \rv}{\p t} - v_0 \uv \right)^2 -\f{1}{4 D_r} \int_0^\t dt \left( \f{\p \uv}{\p t} \right)^2}. 
\eea
Denoting a length  segment of the trajectory by $v_0 dt = dl$, one obtains $\p \rv/\p t = v_0 (\p \rv/\p l)$ and $\p \uv/\p t = v_0  (\p \uv/\p l)$ to get
\bea
{\cal P} [\{\rv (l), \uv (l)\}] \propto e^{- \f{v_0}{4D} \int_0^L dl \left( \f{\p \rv}{\p l} - \uv \right)^2  - \f{v_0}{4 D_r} \int_0^L dl \left( \f{\p \uv}{\p l} \right)^2},
\eea
where  $L=v_0 \t$ is the total length traversed.
This action for the path probability distribution can be written as ${\cal P} [\{ \rv (l), \uv (l) \} ] \propto e^{-\be \cal{E}}$, where  now $\cal{E}$ can be interpreted as the energy of a polymer configuration, and given by   
\bea
\be \mathcal{E}  = \f{A}{2} \int_0^L dl \left( \f{\p \rv}{\p l} - \uv (l) \right)^2 + \f{\k}{2} \int_0^L dl \left(  \f{\p \uv}{\p l} \right)^2,
\label{eq:H}
\eea
where, $\be = 1/\kb T$, and we identify 
$A = v_0/2 D$, and $\k = v_0/2 D_r $.  
This is the energy cost  of a polymer configuration described by $\{\rv(l), \uv(l)\}$, where note that $\rv(l)$ and $\uv(l)$ are independent fields.
In the limit of vanishing translational diffusion ($A \to \infty$), we  require $\dot \rv = v_0 \uv(t)$ and so in this case we can identify $\uv(l) = \p \rv/\p l$ as the local unit tangent vector on the polymer configuration obeying the constraint $|\p \rv/\p l |^2=1$.  
Thus, in this limit, the polymer is effectively described by the second term in the energy expression in  Eq.(\ref{eq:H}) and this corresponds precisely to the worm-like chain (WLC) model with~\cite{Dhar2002,Doi1986} 
\bea
\be \mathcal{E} = \f{\k}{2} \int_0^L dl \left(  \f{\p \uv}{\p l} \right)^2.
\label{eq:wlc}
\eea  
On the ther hand, the limit  $\kappa \to \infty$ requires that $\hat{u}$ be a constant unit vector. Using this input in the first term of Eq.(\ref{eq:H}) leads to the energy functional 
\bea
\be \mathcal{E} = \f{A}{2} \int_0^L dl \left( \f{\p \rv}{\p l} \right)^2- A[\rv(L)-\rv(0)] \cdot {\uv} 
\label{eq:Gauss}
\eea
which corresponds to a Gaussian polymer with a force along the direction ${\uv}$~\cite{Doi1986}.

{In Fig.~\ref{fig:conf} we show a comparison between ($a$)\,a typical configuration of the polymer mapping of the ABP model, ($b$)\,its constant bond-length limit of the WLC model, 
($c$)\,its limit of the Gaussian chain under directed external force for the same parameter values. In Fig.~\ref{fig:conf}($d$)\, we show a related configuration of an extensible semiflexible chain (ESC), discussed in Sec.~\ref{sec:esc}.} 

 To extract equilibrium properties of polymers a common strategy is to perform either Monte-Carlo simulations or Langevin dynamics. In the following section we compare results from such simulations with those obtained from simulations of the ABP dynamics, using the exact mapping of polymer configurations and ABP trajectories. 

\begin{figure}[t]
\begin{center}
\includegraphics[width=8cm]{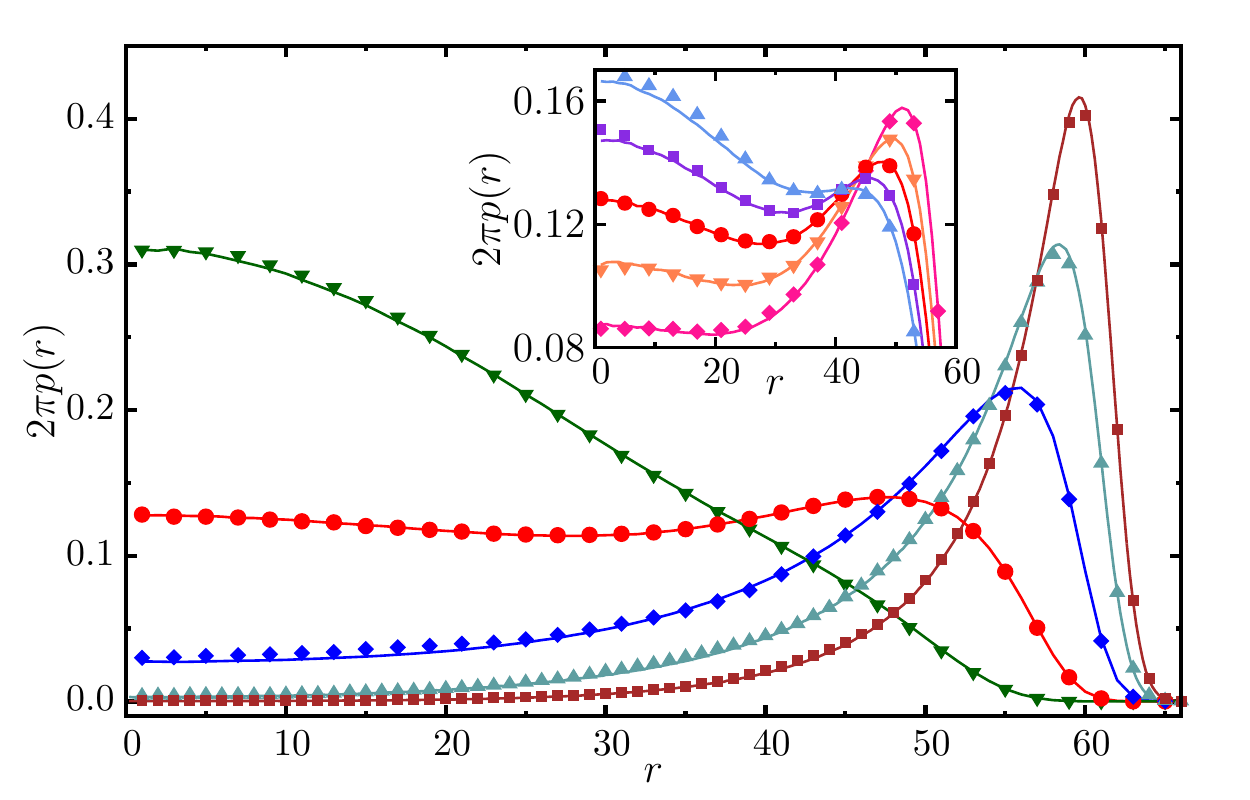}  
\caption{ (color online) Comparison between probability distributions of ABP displacement (solid lines) and end-to-end separation of polymers (points): Here $v_0=1.8\, \s /\t_u$ and $D=0.02 \, \s^2/\t_u$ are held constant. In all the simulations, the first step in the ABP model, and the bond orientation of one end of the polymer are held fixed along the $x$-direction. 
The bimodal distributions corresponding to { $D_r \t_u=0.02\,(\Box)$, $0.03\,(\triangle)$, $0.05\,(\diamond)$, $0.1 \,(\circ)$, $0.2\,(\triangledown)$}. Inset: The same comparisons at $D_r \t_u=0.08\,(\diamond)$, $0.09 \,(\triangledown)$, $0.1\,(\circ)$, $0.11\,(\Box)$, $0.12\,(\triangle)$. 
} 
\label{fig:eeDr}
\end{center}
\end{figure}

\subsection{Comparisons between results from polymer simulations and ABP dynamics using the exact mapping}
Let us now test the polymer mapping numerically by comparing displacement distributions of the ABP with the end-to-end distributions of the mapped polymer.  
We present results in two dimensions ($2d$). Replacing the orientation field $\uv(l) = (\cos \h (l), \sin \h (l)\,)$, the second term in the expression of energy in Eq.(\ref{eq:H}) simplifies to 
$ \f{\k}{2} \int_0^L dl \left(  \f{\p \h}{\p l} \right)^2$. 
After discretization the energy becomes, 
\bea
\be {\cal E} = \sum_{i=1}^{N-1} \f{A}{2 \s} \left[ \bv_i - \s \uv_i  \right]^2 + \sum_{i=1}^{N-1} \f{\k}{2 \s} \left[ \h_{i+1} - \h_i  \right]^2, \label{poly_abp}
\eea
where $\bv_i= \rv_{i+1} - \rv_{i}$ is the bond vector between the $i$-th and $(i+1)$-th bead, and in the first term the vector $\uv_i = (\cos \h_i, \sin \h_i)$. 
The continuum limit is obtained as $\s \to 0$ with $L\equiv v_0 \t =(N-1)\s$, $A/\s$ and $\k/\s$ held constant.
 To perform equilibrium simulations of the polymer, we use the over-damped Langevin equations of motion  
\begin{align}
 \g  \, \dot \rv_i &= - \p {\cal E} /\p \rv_i + \sqrt{2 \g \kb T}\, {\bf F}_i \nn\\
\g_r \, \dot \h_i &= - \p {\cal E} /\p \h_i + \sqrt{2 \g_r \kb T}\,  \L_i,
\end{align}
 where $ {\bf F}_i$ and $\L_i$ denote uni-deviate  Gaussian white noise terms. 
Here $\kb T$ and $\s$ set the unit of energy and length respectively, and $\t_u = \g \s^2/\kb T$ sets the characteristic time over which a bead diffuses over its size $\s$. 
In our simulations we choose the Langevin heat bath characterized by an isotropic friction $\g=\g_r=1/\t_u$. 
The simulations are performed using Euler-Maruyama integration of these equations, with time step $\d t = 0.001 \t_u$. 
 
We perform polymer simulations with $64$ beads, and compare the results with ABP trajectories generated over $L\equiv v_0 \t = 63 \s$. 
We obtain the end-to-end distribution function $p(r)$ for the polymer mapping, and compare the results with probability distributions of the particle-displacements
obtained from the original ABP model. The distributions are normalized to $\int_0^{\infty} p(r)\, 2\pi r \, dr =1$. Three parameters in the ABP model, $D$, $D_r$ and
$v_0$ control the dynamics.

\begin{figure}[t]
\begin{center}
\includegraphics[width=8cm]{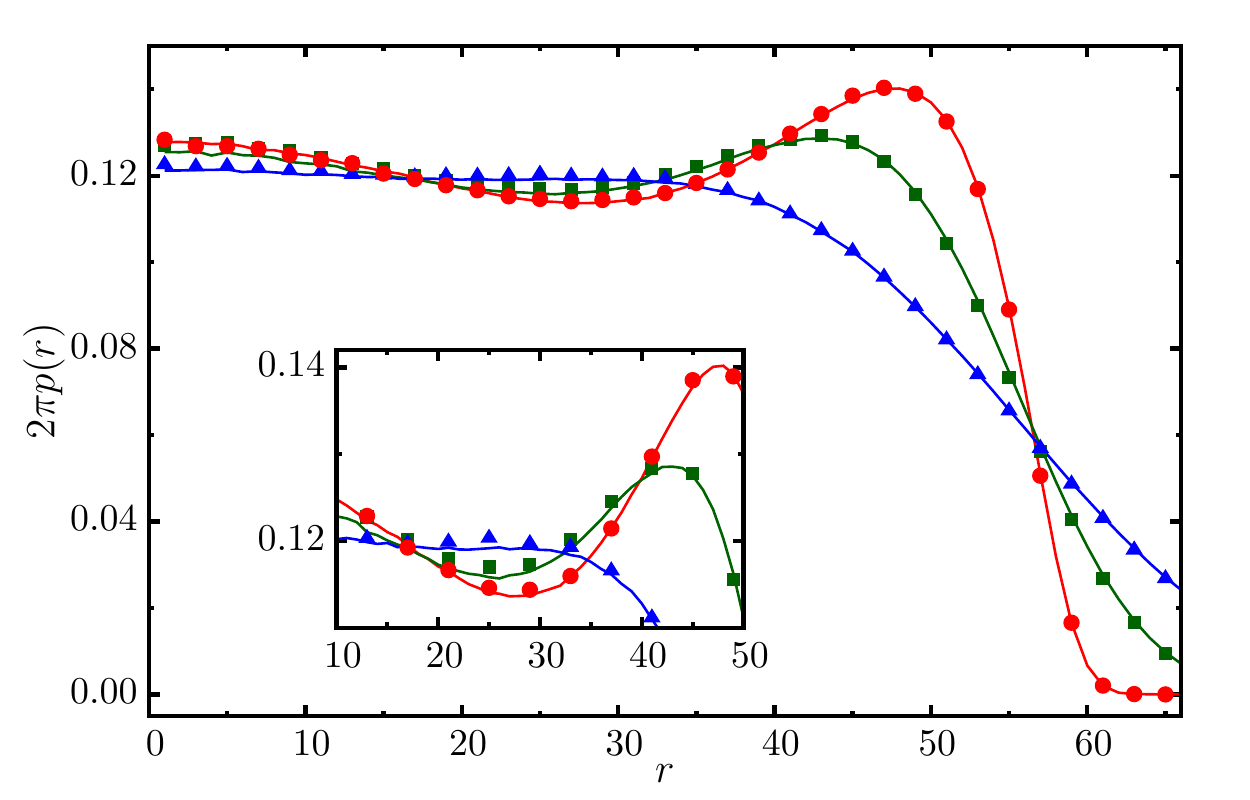}  
\caption{ (color online) Comparison between probability distributions of ABP displacement (solid lines) and end-to-end separation of polymers (points): We keep $v_{o} = 1.8\,\s/\t_u$, $D_r =0.1\, \t_u^{-1}$ for ABP constant, and show the mapping of ABP  to polymer for $D \t_u/\s^2=0.02 \,(\circ)$, $0.5\,(\Box)$, $2\,(\triangle)$.  Inset: The loss of bimodality magnified.} 
\label{fig:eeD}
\end{center}
\end{figure}
 
In Fig.\ref{fig:eeDr} we fix $v_0=1.8\,\s/\t_u$, $D=0.02\,\s^2/\t_u$ and vary $D_r$ of the ABP that maps to different $\k=v_0/2 D_r$ of the semiflexible chain, and keeps the bond stiffness $A=v_0/2 D$ constant. 
The semiflexibility of the chain is determined by the rigidity parameter $L/\ell_p$, the ratio of polymer length $L$ to persistence length $\ell_p= 2\k/(d-1)$. 
In terms of the ABP model, $L/\ell_p = (d-1) D_r \t$.   Fig.\ref{fig:eeDr}  shows the distribution functions in the range of { $0.7 \lesssim L/\ell_p \lesssim 7$}. The inset of 
 Fig.\ref{fig:eeDr}  focuses on the region of bimodality { $2.8 \lesssim L/\ell_p \lesssim 4.2$} recapturing the WLC behavior~\cite{Dhar2002}. 
 The agreement between the two data sets of the ABP model and its polymer mapping is evident from the figure. 
 The relatively large value of $A\,(= v_0/2D = 45 \s^{-1})$ in the corresponding polymer model, for the parameter choice in Fig.~\ref{fig:eeDr}, ensures small bond length fluctuations (within $7.5\%$), allowing to recapitulate the behavior of WLC polymer.

 The bimodality in the distribution function for the ABP means that some of the trajectories will generate small displacements, while some other will produce large displacements. The corresponding polymer will fluctuate between configurations having low to high end-to-end separation. The free energy $F(r,L) = - k_B T \ln[p(r,L)]$ will show a double minima suggesting a non-monotonic force-extension exemplifying a region of negative response in the Helmholtz ensemble~\cite{Dhar2002}.

 \begin{figure}[t]
\begin{center}
\includegraphics[width=8cm]{{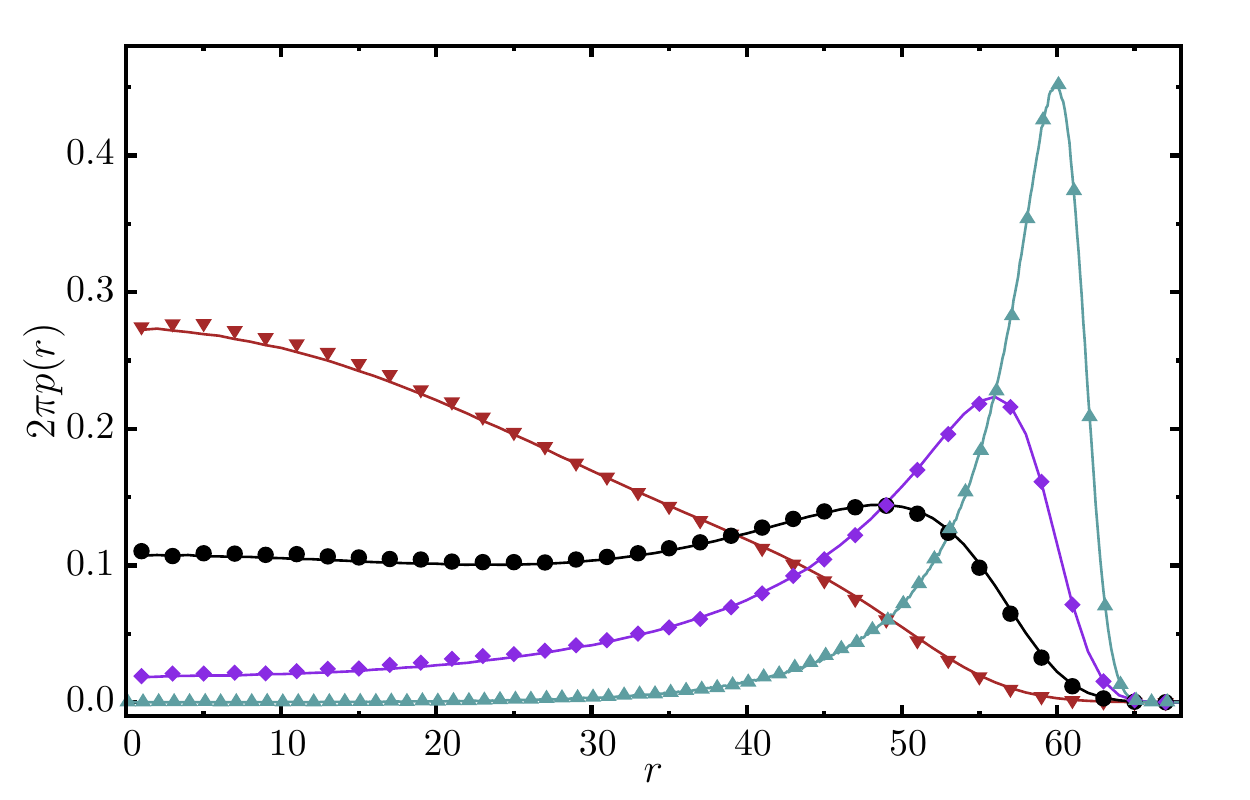}}
\caption{ (color online) Comparison between probability distributions of ABP displacement (solid lines) and end-to-end separation of polymers (points): We kept $D= 1.0\, \s^2/\t_u$ and $D_r=1.0 \t_u^{-1}$ constant and varied $v_0$  keeping $v_0 \t$ constant. 
The data denotes  $v_0 \t_u/\s =10\, (\triangledown)$, $20\,(\circ)$, $40\, (\diamond)$, $100\,(\triangle)$. 
} 
\label{fig:eev0}
\end{center}
\end{figure}

 In  Fig.~\ref{fig:eeD}, we hold $v_0=1.8\,\s/\t_u$, $D_r=0.1\, \t_u^{-1}$ fixed such that at $D=0.02\, \s^2/\t_u$ one obtains clean bimodal distribution as in  Fig.~\ref{fig:eeDr}. We proceed to increase $D$ and
 examine the robustness of the bimodality. At larger $D$, the effective spring constant of the bond lengths $A=v_0/2D$ reduces. Corresponding to 
 $D \t_u/\s^2=0.02, \, 0.5, \, 2$ one finds  two orders of magnitude reduction of spring constant $A \s=45, \, 1.8,\, 0.45$ respectively. This allows large bond- length fluctuations.  Fig.~\ref{fig:eeD} shows  clear numerical agreement between the distribution functions obtained from the two models, exemplifying the mapping. Clearly with reducing $A$ first the contrast of the bimodality reduces as the distribution gets flatter, and finally at $A \s=0.45$ the bimodal structure vanishes. 

Finally, in Fig.~\ref{fig:eev0} we demonstrate  the change in the probability distribution of finite time trajectories of ABP as the propulsion velocity $v_0$ is varied,  keeping $L=v_0 \t$ constant.   
Here we fix the values of $D_r$ and $D$. Increasing $v_0$ increases both $A=v_0/2D$ and $\k = v_0/2 D_r$  together in the polymer mapping. 
At small $v_0$, the  chain remains in the flexible Gaussian regime. With increasing $v_0$ the bond- length fluctuations decrease as the corresponding spring constant $A$ increases. The 
persistence length $\ell_p$ increases as well.  
This leads  the chain towards the WLC regime showing the emergence of bimodality near $v_0 = 20\, \s/\t_u$. 
The comparison of results between the two models show good agreement.

Having established the mapping of the ABP model to the polymer model, the results presented in the following will be interpreted interchangeably. For a fixed $v_0$, the evolution time $\t$ in ABP will be understood in terms of contour length $L=v_0 \t$. 

\begin{figure}[!t]
\begin{center}
\includegraphics[width=1\linewidth]{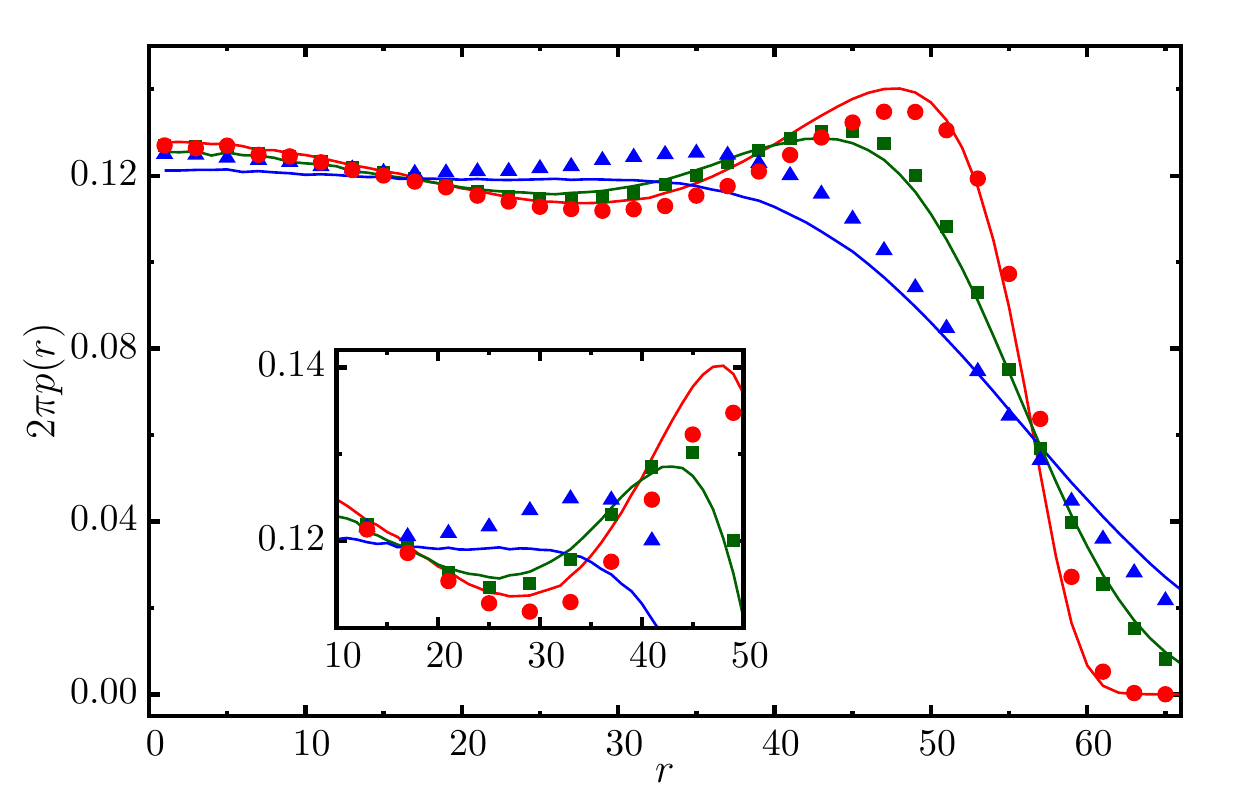} 
\caption{ (color online) Comparison between the ABP model and the extensible semiflexible chain (ESC). We keep parameters $v_0 = 1.8\, \s/\t_u,\, D_r = 0.1\, \t_u^{-1}$ constant. 
The lines show results from the ABP model at $ D \t_u/\s^2 = 0.02$ (red), $0.5$ (green), $2$ (blue).  
Points denote results for ESC at the corresponding $A \s = 45$ (red $\circ$), $1.8$ (green $\Box$), $0.45$ (blue $\triangle$).
Inset: magnified view highlights difference between the results of the two models.
} 
\label{fig:Pee_wlc}
\end{center}
\end{figure}

\subsection{Comparison with polymer model with bending and stretching energy}
\label{sec:esc}
 We briefly comment on how the polymer model described by Eq.\eqref{eq:H} differs from polymers with both bending and stretching energy terms, which we refer to as a extensible semi-flexible chain (ESC).
 Let us consider a polymer with monomer positions ${\bf r}_1,{\bf r}_2,\ldots,{\bf r}_N$. We define bond vectors ${\bf b}_n={\bf r}_{n+1}-{\bf r}_n$, for $n=1,2,\ldots,N-1$ and  let the local tangent ${\bf t}_n={\bf b}_n/b_n$, where $b_n=|{\bf b}_n|$. Then the following energy describes the ESC with  stretching and bending energy terms: 
\begin{align}
\be \mathcal{E}_{\rm ESC} &=\sum_{n=1}^{N-1} \frac{A}{2 \s} (b_n-\s)^2 + \sum_{n=1}^{N-2} \frac{\k}{2 \s} ({\bf t}_{n+1}-{\bf t}_n)^2. 
\label{eq:esc}
\end{align}
In the continuum limit this gives
 \begin{align}
\be \mathcal{E}_{\rm ESC}&=\sum_{n=1}^{N-1} \frac{A}{2 \s} ({\bf r}_{n+1}-{\bf r}_n-\s {\bf t}_n)^2 + \sum_{n=1}^{N-2} \frac{\k}{2 \s} ({\bf t}_{n+1}-{\bf t}_n)^2 ,  \nn\\
&= \int_0^N dn \left[ \frac{A}{2 \s} \left( \f{\partial {\bf r}(n)}{\partial n} -b {\bf t}(n) \right)^2 +  \frac{\k}{2 \s} \left(\f{\partial{\bf t}(n)}{\partial n}\right)^2 \right]~, \nn\\
&= \int_0^L d l  \left[ \frac{A}{2} \left( \f{\partial {\bf r}(l)}{\partial l} - {\bf t}(l) \right)^2 +  \frac{\k}{2} \left( \f{\partial{\bf t}(l)}{\partial l}\right)^2\right]~, \label{eq:Ham2}
\end{align}
where a contour segment is denoted by $l=nb$ and the chain length $L=Nb$. This energy has the same form as in Eq.(\ref{eq:H}),
however,  note that ${\bf r}(l)$ and ${\bf t}(l)$ are {\emph{not independent fields and are related through the equality ${\bf t}(l)= (\partial {\bf r}(l)/\partial l)/|\partial {\bf r}(l)/\partial l |$}}.  On the other hand, ${\bf r}(l)$ and ${\bf u}(l)$ in Eq.\eqref{eq:H}  \emph{are independent} fields.  
Hence, while superficially the two energies in Eq.\eqref{eq:H} and Eq.\eqref{eq:Ham2} look identical, the polymer representation of the active particle needs a different physical interpretation. 
This in fact corresponds to a flexible Gaussian polymer sitting on top of another semiflexible polymer with an inter-polymer interaction that tries to align the two polymers.

In  Fig.~\ref{fig:Pee_wlc}  we compare numerical simulation results from the polymer models in Eq.(\ref{eq:esc}) and Eq.(\ref{poly_abp}). It is clear from the figure that the distributions obtained from the two models are different.
The end-to-end separations they predict do not agree, except in the limit of large $A$. The difference is due to the absence of the constraint  ${\bf t}(l)= (\partial {\bf r}(l)/\partial l)/|\partial {\bf r}(l)/\partial l |$, which is an integral part of ESC,  in the polymer mapping of ABP. For large spring constant $A$, the bond length fluctuations become negligible reducing the polymer configurations corresponding to both the models equivalent to the WLC polymer.

\section{Exact computation of moments for ABP}
\label{sec:moments}

The probability distribution $P(\rv,\uv,t)$ of the position $\rv$ and the active orientation $\uv$,  controlling the self-propulsion velocity $\vv(t) = v_0 \uv(t)$, of the ABP follows the Fokker-Planck  equation
\bea
\p_t P(\rv, \uv, t) = D \nabla^2 P   + D_r \nabla_u^2 P - v_0\, \uv\cdot \nabla P,  \nn
\eea
where $\nabla^2$ is the $d$-dimensional Laplacian operator, and $\nabla_u^2$ denotes the Laplacian in the ($d-1$) dimensional orientation space. We note that the spherical Laplacian can be expressed in terms of the cartesian coordinates ${\bf y}$, defined through $u_i=y_i/y$ where $y=|{\bf y}|$, as $\nabla^2_u= y^2\sum_{i=1}^d \partial^2_{y_i} - [y^2\partial^2_y + (d-1) y \partial_y]$. 
This equation can be derived using the standard procedure of  determining the mean and variance of infinitesimal displacements  in position $\rv(t)$ and orientation $\uv(t)$. We used the {\em Ito} interpretation of the stochastic dynamics. 
The first and last terms on the right hand side describe the translational diffusion and active drift respectively. 
The second term describes orientational diffusion and follows from the result obtained in Eq.(\ref{eq:rot}).

Using the Laplace transform $\tilde P(\rv, \uv, s) = \int_0^\infty dt e^{-s t} P(\rv, \uv, t) $, the Fokker-Planck  equation can be recast in the form,
\bea
-P(\rv, \uv, 0) + s \tilde P(\rv, \uv, s) = D \nabla^2 \tilde P + D_r \nabla_u^2 \tilde P - v_0\, \uv\cdot \nabla \tilde P.  \nn
\eea
Let us define the mean of an arbitrary observable in the Laplace space by $\la \psi \ra_s = \int d\rv \, d\uv\, \psi(\rv, \uv ) \tilde P(\rv, \uv, s)$. 
Multiplying the above equation by $\psi(\rv, \uv)$ and integrating over all possible $(\rv, \uv)$  we find
\bea
-\la \psi \ra_0 + s \la \psi \ra_s = D \la \nabla^2 \psi \ra_s + D_r \la \nabla_u^2 \psi \ra_s + v_0\, \la \uv\cdot \nabla \psi \ra_s,\nn\\
\label{moment}
\eea
where the initial condition sets $\la \psi \ra_0 = \int d\rv \, d\uv\, \psi(\rv, \uv) P(\rv, \uv, 0)$. Without any loss of generality, we consider $P(\rv, \uv, 0) = \d(\rv)\d(\uv - \uv_0)$. 
Eq.(\ref{moment}) can be utilized to compute all the moments as a function of time. In the following, we illustrate the approach by explicitly deriving some of these moments and using them to analyze  the ABP motion (equivalently the polymer model).

\subsection{Orientational correlation}
Let us first consider the evolution of velocity $\vv(t) = v_0 \uv(t)$. Thus we consider $\psi(\rv, \uv) = \uv$. 
It is easy to see that $\la \psi\ra_0 = \uv_0$, $\la \nabla^2 \psi \ra_s =0$,  $\la \uv \cdot \nabla \psi \ra_s = 0$, and
$  \nabla_u^2 \uv = - (d-1) \uv$.
As a result Eq.(\ref{moment}) leads to 
\bea
\la \uv \ra_s = \f{\uv_0}{s+(d-1)D_r}, \nn
\eea
which, after performing inverse Laplace transform gives an exponential decay
\bea
\la \uv (t) \ra = \uv_0 e^{-(d-1)D_r t}.
\eea
From the above derivation, it is easy to see that 
\bea
\la \uv \cdot \uv_0 \ra (t) = e^{-(d-1)D_r t},
\label{eq:uu}
\eea
if one considered $\psi(\rv, \uv) = \uv \cdot \uv_0$. This shows that the orientational correlation decays   
with a correlation time $\t_r = [(d-1) D_r]^{-1}$  in $d$-dimensions. 

The persistence time $\t_r$ sets the unit of time in the problem. Using it along with the 
translational diffusion constant $D$,  the unit of length can be set by $\bar\ell = \sqrt{D/D_r}$ resulting in a unit of velocity $\bar v = \bar\ell/\t_r = \sqrt{D D_r}$. The 
dimension-less activity can be expressed as $\l = v_0/\bar v$.

\begin{figure}[!t]
\begin{center}
\includegraphics[width=8cm]{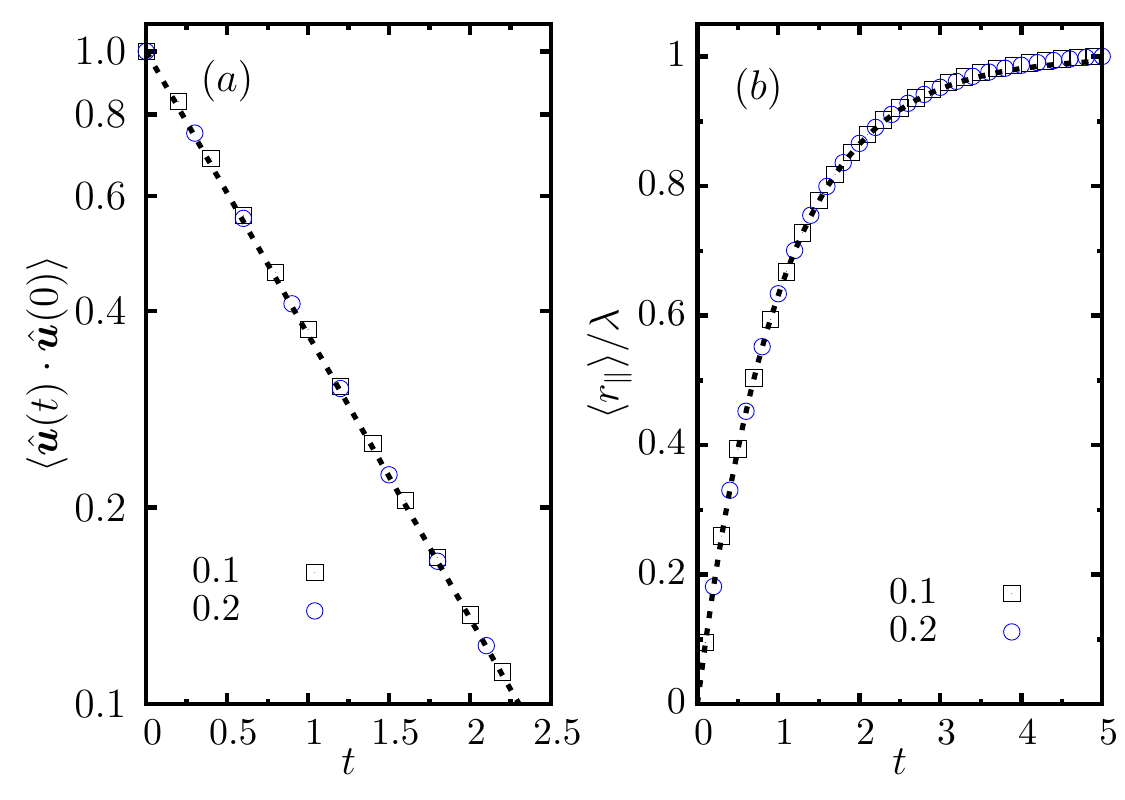}  
\caption{ (color online) ($a$)\,Orientational correlation $\la\uv(t)\cdot\uv(0)\ra$, and ($b$)\,the displacement $\la \rpl \ra$ along the initial orientation $\uv_0$ of the ABP are shown as a function of time $t$ in 2d. Here $D = 1.0\,\s^2/\t_u$, and $v_0 = 1.0 \,\s/\t_u$ are held constant, and we use $D_r \t_u=0.1$, $0.2$. 
The results of numerical simulations are shown by points, and analytic predictions by dashed lines.  
In this figure, and all other figures presented in this section, the length and time axes are expressed in units of $\bar \ell = \sqrt{D/D_r}$ and $\t_r=1/D_r$, respectively. 
The dashed line in ($a$) shows $\la \uv(t) \cdot \uv(0) \ra = e^{-t} $ in the semi-log plot, and in ($b$) shows $\la \rpl \ra\,/\l =  (1-e^{-t})$ with $\l = v_0/\sqrt{D D_r}$.
   } 
\label{fig:ucorr}
\end{center}
\end{figure}

In Fig.\ref{fig:ucorr}($a$) we show simulation results of two-time orientational correlation $\la \uv(t) \cdot \uv(0) \ra$, and its comparison with 
the analytical form $e^{-t}$ where $t$ is expressed in units of $\t_r=1/D_r$ in $2d$.

A mapping of the orientational correlation to the tangent-tangent correlation of the WLC model is possible,  considering the trajectory length $l=v_0 t$ as a polymer segment of the same length. 
The correlation $\la \uv(t) \cdot \uv(0) \ra = \exp(-t/\t_r)$ is then equivalent to $\la \uv(l)\cdot \uv(0)\ra = \exp(-l/\ell_p)$ with $\ell_p = v_0/(d-1)D_r$.   
This is consistent with the WLC result $\ell_p = 2 \k/(d-1)$ and the mapping $\k = v_0/2D_r$ between the ABP and its corresponding polymer model.

\subsection{Displacement}
Using $\psi = \rv$ in Eq.(\ref{moment}), along with the result $\la \uv \ra_s = \uv_0/(s+(d-1)D_r)$ allows us to obtain 
\bea
\la \rv \ra_s = \f{v_0 \, \uv_0}{s (s+(d-1)D_r)},
\label{eq:rpl}
\eea
which leads to
\bea
\la \rv \ra (t) = \f{v_0 \, \uv_0}{(d-1) D_r} \left( 1 - e^{-(d-1)D_r\, t} \right).
\label{eq:rv}
\eea
 Let us define the displacement components along and perpendicular to the initial orientation as
\begin{align}
\bm{\rpl}= (\rv.\uv_0) \uv_0,~~~~\rp = \rv-\bm{\rpl}~. \label{eq:rplrp}
\end{align}
We then see that the mean displacement
along the initial orientation $\la \rpl \ra = \la \rv \cdot \uv_0 \ra$ grows and saturates to a finite value as $\la \rpl \ra /\bar \ell= \l (1-e^{-t/\t_r})$, where $\l = v_0/\bar v$ is the dimensionless parameter
controlling activity~(see Fig.\ref{fig:ucorr}($b$)\,). On the other hand,  the average displacement vector perpendicular to $\uv_0$ vanishes $\la \rp \ra = 0$.   

  \begin{figure}[!t]
\begin{center}
\includegraphics[width=8cm]{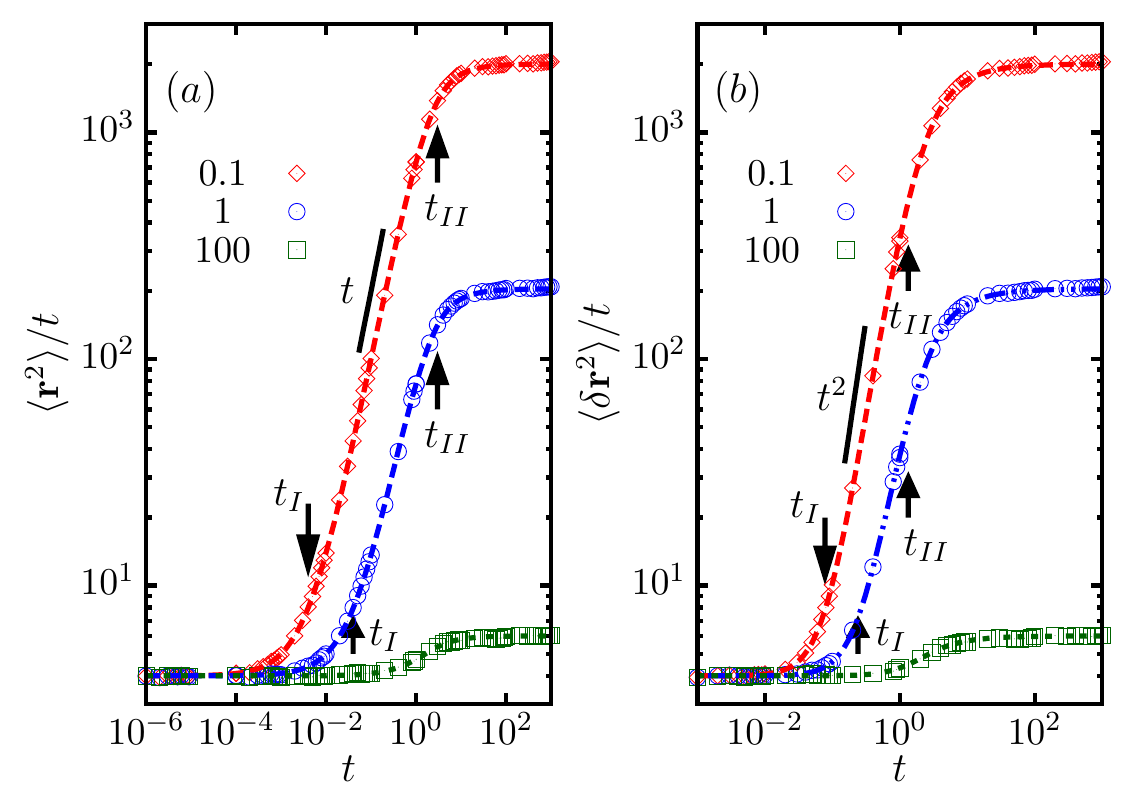}  
\caption{ (color online) Time dependence of ratios $\la r^2\ra/t$ and $\la \d r^2 \ra/t$ in 2d.   
$D_{r} = 1.0\,\t_u^{-1}$, $v_0=10\,\s/\t_u$ 
are held constant. The  results of numerical simulation at $D \t_u/\s^2=0.1,\, 1,\, 100$   
are shown by points denoted in the two figures.
 Dashed lines show plots of $\la \rv^2 \ra$ and  $\la \d \rv^2 \ra$  obtained from Eq.(\ref{eq:rpl}) and (\ref{eq:r2}). 
For individual curves, arrows denote $t_I= (2d/\l^2) \t_r$, $t_{II} = [3/(d-1)]\t_r$  in ($a$), and $t_I = [3d/(d-1)]^{1/2} \t_r$ and $t_{II} = [4/3(d-1)] \t_r$ in ($b$), with $d=2$.
 } 
\label{fig:r_squared}
\end{center}
\end{figure}

\subsection{Fluctuations of the displacement}
 Let us now consider $\psi(\rv, \uv) = \rv^2$ and calculate the time dependence of $\la \rv^2 \ra(t)$. It is easy to see that $\la \psi \ra_0 = 0$ and $\la \nabla_u^2 \psi \ra_s =0$. The average $\la \nabla^2 \rv^2 \ra_s = 2 d\la 1 \ra_s$. Note that $\la 1 \ra_s = \int d\rv d\uv \tilde P = \int d\rv d\uv \int_0^\infty dt e^{-st} P  = \int_0^\infty dt e^{-st} \{  d\rv d\uv P\} = \int_0^\infty dt e^{-st}  = 1/s$.  Further, $\la\uv\cdot \nabla \rv^2 \ra_s = 2 \la \uv \cdot \rv \ra_s$. Thus Eq.\,(\ref{moment}) leads to $s \la \rv^2 \ra_s = 2d D/s \, + 2 v_0 \la \uv \cdot \rv \ra_s$. To complete the calculation, one needs to evaluate $\la \uv \cdot \rv \ra_s $ using the same  Eq.\,(\ref{moment}). One may proceed like before, utilizing the relation $\nabla_u^2 \uv = -(d-1)\uv$, $\la \uv \cdot \nabla \psi \ra_s = \la \uv^2 \ra_s = 1/s$, to get $s \la \uv \cdot \rv \ra_s = -(d-1) D_r  \la \uv \cdot \rv \ra_s +v_0/s$. This gives
\bea
\la \uv \cdot \rv\ra_s = \frac{v_0}{s(s+(d-1)D_r)},
\label{eq:urs}
\eea 
which leads to the cross-correlation
\bea
\la \uv \cdot \rv\ra = \f{v_0}{(d-1)D_r} \left( 1 - e^{-(d-1)D_r\, t} \right). \nn
\label{eq:vr}
\eea
Plugging the relation from Eq.\eqref{eq:urs} into Eq.(\ref{moment}) one finds 
\bea
\la \rv^2\ra_s =  \f{2d D}{s^2} + \f{2 v_0^2}{s^2(s+(d-1)D_r)}. 
\label{eq:r2s}
\eea   
Performing the inverse Laplace transform, we obtain  
\bea
\la \rv^2 \ra &=& 2 d \left( D + \f{v_0^2}{(d-1) d D_r}\right) t \nn\\
&-& \f{2v_0^2}{(d-1)^2 D_r^2 } \left(1 -e^{-(d-1)D_r\, t} \right). 
\label{eq:r2}
\eea
In the limit of $t \ll \t_r = 1/(d-1) D_r$, 
the motion is dominated by the simple translational diffusion, $\la \rv^2 \ra \approx 2dD\,t$.
In the long time limit, 
the equation gives diffusive scaling $\la \rv^2 \ra = 2 d\,D_{\rm eff} t$ with the effective diffusion constant in $d$-dimensions
\bea
D_{\rm eff} = D  +  \f{v_0^2}{(d-1) d D_r}.
\eea
A series expansion of Eq.(\ref{eq:r2}) around $t=0$ gives
\bea
\la \rv^2 \ra = 2 d D t + v_0^2 t^2 - \f{1}{3} v_0^2(d-1)D_r t^3 +{\cal O}(t^4). \nn
\eea
This shows that $\la \rv^2 \ra$ will crossover from a diffusive $\sim t$  to ballistic $\sim t^2$  scaling at $t_I = (2 d /\l^2) \t_r$.  
This is expected to be followed by another crossover from  ballistic to diffusive behavior near $t_{II} \approx [3 /(d-1)\,] \t_r$.  
These crossovers along with the estimated crossover-points $t_I$ and $t_{II}$ are shown in Fig.~\ref{fig:r_squared}($a$) for an ABP moving in 2d. 
The simulation results agree with the above estimates.
In the limit of $D=0$,  only a single crossover from $\la \rv^2 \ra\sim t^2$ to $\la \rv^2 \ra \sim t$ at $t_{II} D_r \approx 3/(d-1)$ survives.

Using Eq.(\ref{eq:rv}) and (\ref{eq:r2}), one can calculate $\la \d \rv^2\ra = \la \rv^2\ra - \la \rv \ra^2$.  
The lines through the simulation results plotted with points in Fig.~\ref{fig:r_squared}($b$) correspond to this relation. 
In the small time limit it can be expanded to give
\bea
\la \d \rv^2 \ra &=& 2 d D t + \f{2}{3} (d-1) D_r v_0^2 t^3 \nn\\
&& -\f{1}{2} (d-1)^2 D_r^2 v_0^2 t^4 +{\cal O}(t^5). 
\eea
Thus the mean squared displacement $\la \d \rv^2 \ra$ is expected to show crossovers from a diffusive $\sim t$ scaling to $\sim t^3$ scaling 
at $t_I \approx [3d/(d-1)]^{1/2} \t_r/\l$. 
This would be followed by another crossover back to diffusive
scaling near $t_{II} \approx [4/3 (d-1)] \t_r $. 
These crossovers obtained from simulations in 2d and their comparison with the above analyses are shown in Fig.~\ref{fig:r_squared}($b$). 
In the limit of $D=0$, only a single crossover from $\la \d \rv^2 \ra\sim t^3$ to $\la \rv^2 \ra \sim t$ at $t_{II} \approx  [4/3 (d-1)] \t_r $ survives.

The expression for $\la \rv^2 \ra$ in Eq.(\ref{eq:r2}) can  easily be mapped to find the expression for the end-to-end separation for WLC model, setting $D=0$.
As before, we use $\k = v_0/2 D_r$, $l=v_0 t$ to obtain
\bea
\la \rv^2 \ra &=& \f{4 \k l}{d-1} - \f{8 \k^2 (1-e^{-\f{(d-1)l}{2 \k}})}{(d-1)^2},
\eea
a well known result of the WLC model~\cite{Dhar2002}.

\subsection{Components of displacement fluctuations}

Due to the persistence of motion, the fixing of initial active orientation of the ABP leads to asymmetric displacements, characetrized by   
 $\la \rpl^2 \ra$ and $\la \rp^2 \ra$, where $\rpl$ and $\rp \perp \uv_0$ are defined in Eq.\eqref{eq:rplrp}. 
Without any loss of generality, we assume that the initial orientation of activity $\uv_0$ is in  the $x$-direction, 
$\uv_0 = \hat x$. Using $\rpl^2 = x^2$ as $\psi$ in Eq.(\ref{moment}), we get 
\bea
s \la \rpl^2 \ra_s = 2D/s \, + 2v_0 \la x u_x \ra_s. 
\label{eq:rpl_1}
\eea
To proceed, we again consider $\psi = x u_x$ in Eq.(\ref{moment}), giving $\la \psi\ra_0 = 0$, $\nabla^2 \psi = 0$, $\nabla_u^2 \psi = - (d-1) x u_x$, $\uv \cdot \nabla \psi = u_x^2$, to get 
  $s \la x u_x \ra_s = -(d-1) D_r \la x u_x\ra_s + v_0 \la u_x u_x \ra_s$
leading to
\bea 
\la x u_x\ra_s = \f{v_0}{s + (d-1) D_r} \la u_x u_x \ra_s. 
\label{eq:xu}
\eea
At this stage we need to calculate $\la u_x^2 \ra$. 
Using $\la u_x^2  \ra_0 = 1$, $\nabla^2 u_x^2  = 0$, 
$\la \nabla_u^2 u_x^2  \ra_s= - 2 d \la u_x^2\ra_s +2/s$, 
$\uv \cdot \nabla u_x^2  =0$ 
in Eq.(\ref{moment}) we find 
\bea
\la u_x u_x \ra_s = \f{(s+2D_r)}{s(s+2 d D_r)}.
\label{eq:uu}
\eea 
In calculating $\la \nabla_u^2 u_x^2\ra_s$ we used the general relation 
\bea
\nabla_u^2 (u_i u_j) = -2d \, u_i u_j + 2 \d_{ij}.
\label{eq:del_uu}
\eea
To derive this, let us consider $u_i u_j = r_i r_j / r^2$ and $\nabla^2 = \p_r^2 + (1/r^2) \nabla_u^2$. Note that this Laplacian $\nabla^2$ operates on the active orientation, not on the position vector of the particle.  If $f$ is a function of $\uv$ alone, $\nabla^2 f = (1/r^2) \nabla_u^2 f$. It is easy to directly calculate $\nabla^2 (r_i r_j /r^2)$ component- wise in cartesian coordinates. The result $\nabla^2 (r_i r_j /r^2) = -2d\, r_i r_j/r^4 + (2/r^2) \d_{ij} $ then leads to 
Eq.\eqref{eq:del_uu}.

Using Eqs.\,(\ref{eq:rpl_1}), (\ref{eq:xu}) and (\ref{eq:uu}), we obtain 
\bea
\la \rpl^2 \ra_s = \f{2D}{s^2} + \f{2 v_0^2 (s+ 2 D_r)}{s^2(s+ (d-1) D_r)(s+2 d D_r)}. 
\eea
Performing inverse Laplace transform one finds
\bea
\la \rpl^2 \ra &=& 2 \left(D + \f{v_0^2}{(d-1) d D_r} \right) t + \f{v_0^2}{D_r^2} \left(\frac{(d-1) e^{-2 d D_r t}}{d^2 (d+1)} \right.\nn\\
&& \left. +\frac{2 (3-d) e^{-(d-1) D_r t}}{(d-1)^2 (d+1)} +\frac{d^2-4 d+1}{(d-1)^2 d^2} \right).
\eea
\begin{widetext} 
This can be used to calculate the relative fluctuations $\la \d \rpl^2 \ra = \la \rpl^2 \ra -  \la \rpl \ra^2 $ and $\la \d \rp^2 \ra = \la \rp^2 \ra=\la \rv^2 - \rpl^2 \ra$, since $\la \rp \ra=0$. They are given by
\bea
\la \d \rpl^2 \ra &=& 2 \left(D + \f{v_0^2}{(d-1) d D_r} \right) t + \f{v_0^2}{D_r^2} \left(\frac{(d-1) e^{-2 d D_r t}}{d^2 (d+1) } +\frac{8 e^{-(d-1) D_r t}}{(d-1)^2 (d+1) } 
-\f{e^{-2(d-1)D_r t}}{(d-1)^2 } -\frac{4 d-1}{(d-1)^2 d^2 } \right), \label{eq:drpl2}\\
\la \d \rp^2 \ra &=& 2 (d-1)\left( D + \f{v_0^2}{(d-1)d D_r}\right)\,t + \f{v_0^2}{D_r^2} \left( \f{4 e^{-(d-1)D_r t}}{d^2-1} - \f{(d-1)e^{-2d D_r t}}{d^2(d+1)} - \f{3d-1}{d^2(d-1)}\right).
\label{eq:drpp2}
\eea
\end{widetext}

\begin{figure}[!t]
\begin{center}
\includegraphics[width=8cm]{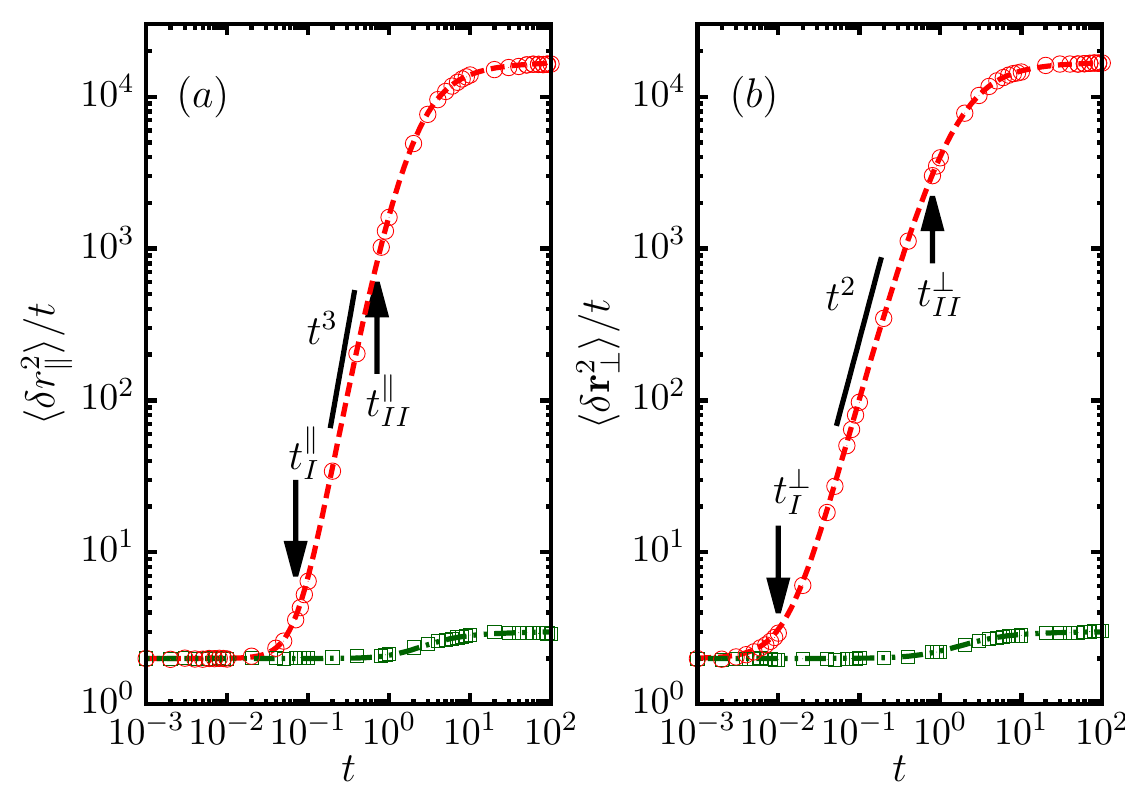}  
\caption{ (color online) The plots of $\la \d \rpl^2 \ra$ and $\la \rp^2 \ra$ as a function of time $t$ in 2d. 
$D_{r} = 1.0\, \t_u^{-1}$, $D=1.0 \,\s^2/\t_u$ are held constant.
The data for $v_0 \t_u/\s =1\,(\square),\,130\,(\circ)$  are shown in the two figures.
The dashed lines show plots of $\la \d \rpl^2 \ra$  and  $\la \d \rp^2 \ra$  obtained from Eq.(\ref{eq:dr2comp1}) and (\ref{eq:dr2comp2}). 
For the expressions of $t^\parallel_{I,II}$ and $t^\perp_{I,II}$, see the discussion after Eq.(\ref{eq:dr2comp_exp}).
The solid lines in the figures denote the intermediate scaling.
} 
\label{fig:msd}
\end{center}
\end{figure}

In two dimensions ($d=2$) the mean squared displacements  
of the parallel and perpendicular components simplify to 
\bea
\la \d \rpl^2 \ra &=& 2 \left(D + \f{v_0^2}{2 D_r} \right) t \nn\\
&&+\f{v_0^2}{D_r^2} \left(\frac{1}{12} e^{-4 D_r t}  -e^{-2 D_r t} +\frac{8}{3} e^{- D_r t} -\frac{7}{4} \right)\nn\\  \label{eq:dr2comp1}\\
\la \d \rp^2 \ra &=& 2 \left( D + \f{v_0^2}{2 D_r}\right)\,t \nn\\
&&+ \f{v_0^2}{D_r^2} \left( - \f{1}{12} e^{-4 D_r t} + \f{4}{3} e^{-D_r t} - \f{5}{4}\right).
\label{eq:dr2comp2}
\eea
 For $D=0$, these agree with the results obtained in \cite{Basu2018}.
In Fig.\ref{fig:msd} we show a comparison of these analytic estimates with numerical simulations of ABP model  at two different propulsion velocities $v_0$, in 2d.
The analytic expressions of Eq.(\ref{eq:dr2comp1}) and (\ref{eq:dr2comp2}) are plotted by lines, and the simulation results by points. The results agree with each other. 
In the long time limit, both the components, $\la \d \rpl^2 \ra$ and $\la \d \rp^2 \ra$,  show the same diffusive scaling $\sim t$. However, at shorter time their respective behaviors differ. 
We can further use the analytic expressions
to extract the observed crossovers in the dynamics of Fig.\ref{fig:msd}.
Performing an expansion around $t=0$  in $2d$ we find,
\bea
\la \d \rpl^2 \ra &=& 2 D t + \f{1}{3} v_0^2 D_r^2 t^4 - \f{7}{15} v_0^2 D_r^3 t^5 + {\cal O}(t^6)\nn\\
\la \d \rp^2 \ra &=& 2 D t + \f{2}{3} v_0^2 D_r t^3 - \f{5}{6} v_0^2 D_r^2 t^4 + {\cal O}(t^5).
\label{eq:dr2comp_exp}
\eea 
The parallel component $\la \d \rpl^2 \ra$ first crosses over from 
$\sim t$ to $\sim t^4$ at 
$t^\parallel_I = (6/\l^2)^{1/3} \t_r$    
followed by another crossover to $\sim t$ at $t^\parallel_{II} \approx (5/7)\t_r$ independent of the amount of active drive $\l = v_0/\sqrt{D D_r}$.  
Note that both the crossovers will be observable only if $t^\parallel_{II} > t^\parallel_I$, requiring $ \l  > 4.1 $. 

The transverse fluctuations $\la \d \rp^2 \ra$ first crosses over from $\sim t$ to $\sim t^3$ scaling at 
$t^\perp_I = [3/\l^2]^{1/2} \t_r$  
followed by another crossover to $\sim t$ at 
$t^\perp_{II}  \approx (4/5)\t_r$ independent of $\l$. 
These two crossovers will be observable if  $t^\perp_{II} > t^\perp_I$, requiring $ \l > 2.2 $.
We present simulation results in 2d, and their comparison with analytic expressions  in Fig.\ref{fig:msd}.
 In this figure, both the above mentioned conditions are satisfied for $\l=130$, and broken for $\l=1$. As a result, we observe the 
two crossovers only at $\l=130$ in Fig.\ref{fig:msd}($a$) and ($b$). Whereas, the same figures for $\l=1$ shows approximate diffusive scalings all through. The lines through the
simulation results are plots of Eq.s\,(\ref{eq:dr2comp1}) and (\ref{eq:dr2comp2}).  
As is evident from Eq.(\ref{eq:dr2comp_exp}), in the absence of translational diffusion, the short time scaling behaviors are dominated by $\la \d \rpl^2 \ra \sim t^4$ and $\la \d \rp^2 \ra \sim t^3$,  as was already been pointed out in \cite{Basu2018}.

\subsection{Fourth moment}
\begin{figure}[!t]
\begin{center}
\includegraphics[width=8cm]{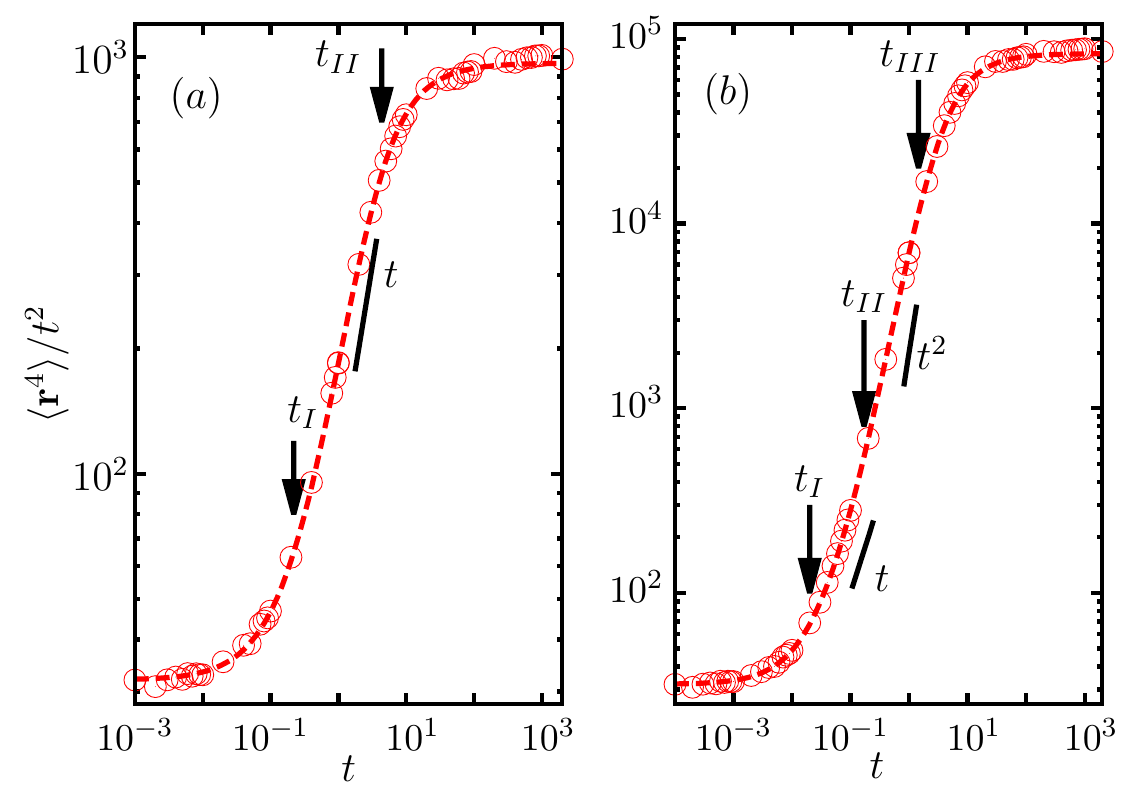} 
\caption{ (color online) Evolution of the ratio $\la r^4 \ra/t^2$ with time in 2d. 
$D_r = 1.0\, \t_u^{-1}$ and $D=1.0\, \s^2/\t_u$ are held constant. The points denote simulation results, and the dashed lines denote analytic prediction in Eq.(\ref{r4d2}). 
$(a)$~$v_0=3\,\s/\t_u$ shows two crossovers at $t_{I} = 0.22\, \t_r$, $t_{II} = 4.36\, \t_r$ , 
$(b)$~$v_0=10\,\s/\t_u$ shows three crossovers at crossover times $t_{I}\approx 0.02\, \t_r$, $t_{II} = 0.17\, \t_r$ and $t_{III} = 1.45\, \t_r$. 
The black solid lines in the two curves indicate the intermediate scaling behaviors.
} 
\label{fig:r4}
\end{center}
\end{figure}

The calculation of $\la \rv^4 \ra$ involves the following steps: 
(i)~$s \la \rv^4 \ra_s = 4 (d+2) D \la \rv^2 \ra_s + 4 v_0 \la (\uv \cdot \rv) \rv^2 \ra_s$, evaluating which requires us to consider the equation  
(ii)~$[s + (d-1)D_r] \la (\uv \cdot \rv) \rv^2  \ra_s = (4 + 2d) D \la \uv \cdot \rv\ra_s + v_0 \la \rv^2 \ra_s + 2 v_0 \la (\uv \cdot \rv)^2 \ra_s$. 
This in turn requires us to consider the equation  
(iii)~$(s + 2 d D_r) \la (\uv \cdot \rv)^2 \ra_s = \f{2D}{s} + 2 D_r \la r^2\ra_s   + 2 v_0 \la \uv \cdot \rv \ra_s $.
Using Eq.(\ref{eq:del_uu}) one can show that $\nabla_u^2 [(\uv \cdot \rv)^2] = 2 r^2 - 2d (\uv \cdot \rv)^2$. 
The expressions for $\la \uv \cdot \rv\ra_s$ and $\la \rv^2\ra_s$ were already evaluated in Eq.(\ref{eq:urs}) and (\ref{eq:r2s}). Thus one can use
all these steps to complete the calculation leading to

\bea
 \la \rv^4 \ra_s &=& \f{8}{s^3} \left[ d(d+2) D^2 + D v_0^2\f{(d+2)(3s + 2 (d-1) D_r)}{(s+(d-1)D_r)^2} \right. \nn\\
&+& \left. v_0^4 \f{3s + 2 (d+2) D_r}{(s+(d-1) D_r)^2 (s + 2d D_r)} \right] .
\label{r4s}
\eea  
Apart from the factor $d(d+2)$ in the first term, this agrees with Eq.~(34) of Ref.~\cite{sevilla2015}. Note that this result is independent of the initial orientation $\uv_0$ and so the difference persists even after averaging over initial conditions. 
Performing the inverse Laplace transform, we obtain the time evolution of the fourth moment in  $d$-dimensions, 
\begin{widetext}
\begin{align}
&\la \rv^4(t) \ra = \frac{4 (d-1) v_0^4 e^{-2 d D_r t}}{d^3 (d+1)^2 D_r^4} 
-\frac{8 \left(d^2 v_0^4+10 d v_0^4+25 v_0^4\right) e^{-(d-1) D_r t}}{(d-1)^4 (d+1)^2 D_r^4}
+\frac{4 \left(d^3 v_0^4+23 d^2 v_0^4-7 d v_0^4+v_0^4\right)}{(d-1)^4 d^3 D_r^4}\nn\\
&+\frac{8 t e^{- (d-1) D_r t } \left(d^3 D D_r v_0^2+2 d^2 D D_r v_0^2-d D D_r v_0^2+d v_0^4-2 D D_r v_0^2-7 v_0^4\right)}{(d-1)^3 (d+1) D_r^3}\nn\\
&   +\frac{4 t^2 \left(d^5 D^2 D_r^2-3 d^3 D^2 D_r^2+2 d^3 D D_r v_0^2+2 d^2 D^2 D_r^2+2 d^2 D D_r v_0^2-4 d D D_r v_0^2+d v_0^4+2 v_0^4\right)}{(d-1)^2 d D_r^2}\nn\\
 &  -\frac{8 t \left(d^4 D D_r v_0^2+d^3 D D_r v_0^2-2 d^2 D D_r v_0^2+d^2 v_0^4+6 d v_0^4-v_0^4\right)}{(d-1)^3 d^2 D_r^3} .
 \label{eq:r4_abp}
\end{align}
\end{widetext}
Again in $d=2$ the relation simplifies to
\begin{widetext}
\begin{align}
& \la \rv^4(t) \ra =\frac{8 t^2 \left(4 D^2 D_r^2+4 D D_r v_0^2+v_0^4\right)}{D_r^2}+\frac{8 t e^{-D_r t} \left(12 D
   D_r v_0^2-5 v_0^4\right)}{3 D_r^3}-\frac{2 t \left(16 D D_r v_0^2+15 v_0^4\right)}{D_r^3} \nn\\
&+\frac{v_0^4 e^{-4 D_r t}}{18 D_r^4}-\frac{392 v_0^4 e^{-D_r t}}{9 D_r^4}+\frac{87 v_0^4}{2 D_r^4}.
\label{r4d2}
\end{align}
\end{widetext} 
 For $D=0$ this agrees with the expression in \cite{Dhar2002} and we have also verified that our result for $d=3$ agrees with \cite{Hermans1952}. 
Eq.(\ref{r4d2}) is plotted by dashed lines in Fig.\ref{fig:r4}. As is clearly seen from the figure, the 2d simulation data (points) agree well with this analytic expression. 
In the limit of $t \gg 1/D_r$, the first term in the above expression dominates to give $\la \rv^4(t) \ra \sim t^2$. 
The change in scaling with $t$ as observed from the figure can be better understood by considering the expansion of the expression in Eq.(\ref{r4d2}) around $t=0$, 
\bea
 \la \rv^4(t) \ra &=& 32 D^2 t^2 + 16 D v_0^2 t^3 + \left(v_0^2 -   \f{16}{3} D D_r  \right) v_0^2 t^4 \nn\\
&& - \f{2}{3} v_0^2 D_r \left(  v_0^2 - 2 D D_r \right) t^5+ {\cal O}(t^6).\nn
\eea 
This relation shows that at smallest time $\la \rv^4(t) \ra \sim t^2$, which crosses over to $ \la \rv^4(t) \ra \sim t^3$  at 
$t_I  = (2/\l^2)\t_r$.  
A second crossover from $\sim t^3$ to $\sim t^4$ may appear at $t_{II} = [48 /(3 \l^2 - 16)\,]\t_r$ provided $\l^2 > 16/3$.
At a longer time,   
$t_{III}  \approx \f{1}{2} \f{3 \l^2 - 16}{\l^2 - 2} \t_r$
the time-dependence is expected to show a third
cross- over back to $\la \rv^4(t) \ra \sim t^2$. 

It is clear that whether all these crossovers will be observable depends on the activity parameter $\l$. For example, the requirement to 
observe the third crossover  $t_{III}> t_{II}$ is satisfied only when $\l^2 > 56/3$. As can be seen from Fig.\ref{fig:r4}($a$), $\la \rv^4 \ra$  shows 
$\sim t^2$ to $\sim t^3$ crossover at $t_I$, and a direct crossover back to $\sim t^2$ beyond $t_{II}$ at $v_0=3\, \s/\t_u$ that obeys the condition $\l^2 < 56/3$.  On the other 
hand Fig.\ref{fig:r4}($b$) at $v_0=10\, \s/\t_u$, satisfying the condition $\l^2 > 56/3$, clearly shows all the three crossovers discussed above. The crossover 
points indicated in the figures correspond to the expressions derived above.

Moreover, the expression for the fourth moment of the persistent walk corresponding to the WLC polymer in $d$-dimensions is easily obtainable by setting $D=0$ in Eq.(\ref{r4s}), 
\bea
\la \rv^4 \ra_s = 8 v_0^4 \f{3s + 2 (d+2) D_r}{s^3 (s+ (d-1)D_r\,)^2\, (s+ 2 d\, D_r\,)}.
\eea
The inverse Laplace transform of this relation gives the evolution, 
\begin{align}
&\la \rv^4(t) \ra = 4 v_0^4 \left(\frac{ (d-1) e^{-2 d D_r t}}{d^3 (d+1)^2 D_r^4}
-\frac{2 \left(d+5\right)^2  e^{-(d-1)D_r t}}{(d-1)^4 (d+1)^2 D_r^4} \right. \nn\\
&  \left. -\frac{2 \left(d^2 +6 d -1\right)  t}{(d-1)^3 d^2 D_r^3} 
+\frac{(d^3 +23 d^2 -7 d +1)}{(d-1)^4 d^3 D_r^4} \right. \nn\\
&  \left. -\frac{2  (7-d) \,t e^{-(d-1) D_r t}}{(d-1)^3 (d+1) D_r^3}+\frac{ (d+2)\, t^2}{(d-1)^2 d D_r^2}\right).
\end{align}
Replacing $\k = v_0/2 D_r$, $l=v_0 t$ provides the well known result for $\la \rv^4 (l) \ra$ of WLC model~\cite{Dhar2002}. 

\section{End-to-end distribution with increasing chain length}
\label{sec:chainL}

\begin{figure}[!t]
\begin{center}
\includegraphics[width=8cm]{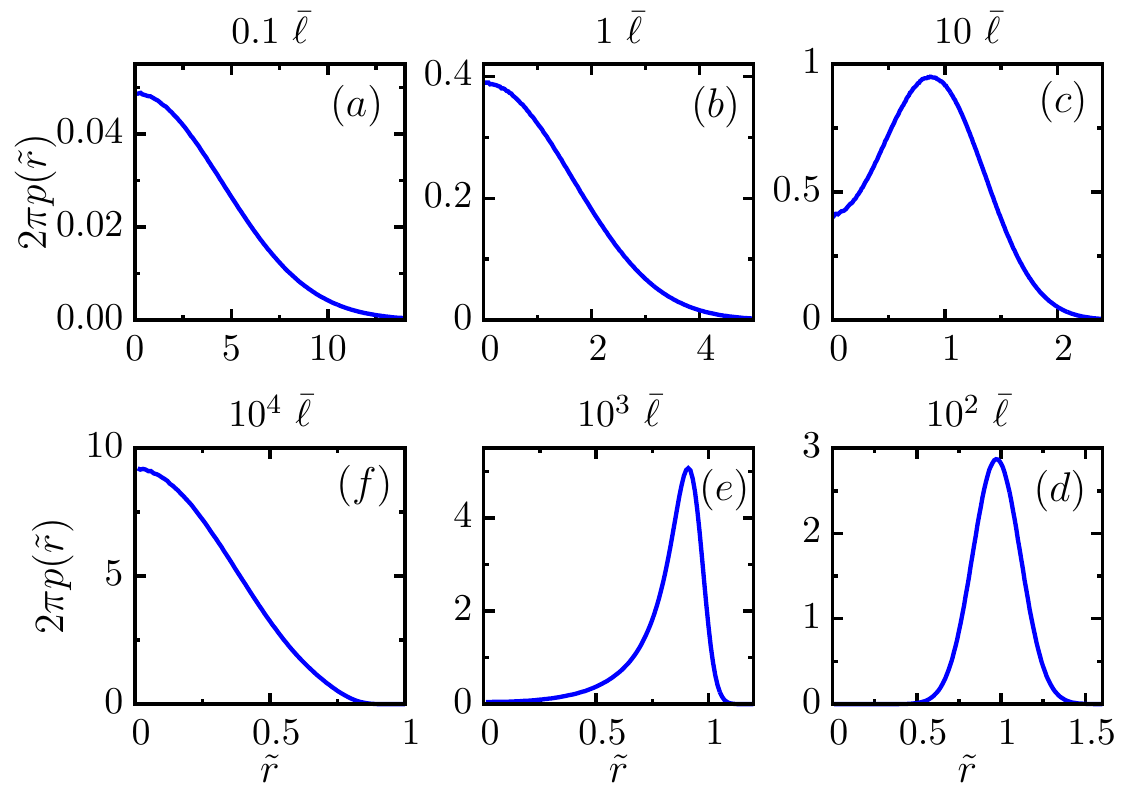}  
\caption{ (color online) The displacement distributions at $D_{r} = 1.0\, \t_u^{-1}$, $v_0=10\, \s/\t_u$ and $D=0.1\, \s^2/\t_u$ over different time-segments indicated by the $v_0 t$ values shown in the figures. 
The persistence length $\ell_p = 10^3 \bar \ell$. The length of trajectories considered are $L =0.1 \,\bar \ell\,(a), 1.0\, \bar\ell\,(b), 10\,\bar\ell\,(c) \dots, 10^4 \bar \ell\,(f)$.
} 
\label{fig:distb1}
\end{center}
\end{figure}

In the equivalent polymer model, the dynamical crossovers with observation time $\t$ translate into similar behavior of the end-to-end separation $\la \rv^2 \ra$ with increasing contour length $L = v_0 \t$ for a polymer
with a given $A=v_0/2D$ and $\k = v_0/2D_r$.   
In Fig.\ref{fig:distb1} we plot the distribution functions $p(\tilde r)$ of the scaled separation $\tilde r = r/L$. For the given choice of parameters, 
$D_{r} = 1.0 \t_u^{-1}$, $v_0=10\, \s/\t_u$ and $D=0.1\, \s^2/\t_u$, 
the persistence length of such a chain in 2d is  $\ell_p = v_0/D_r = 10^3 \, \bar \ell$, where $\bar \ell = \sqrt{D/D_r} = 10^{-2} \s$.
As is clear from Fig.\ref{fig:distb1}($a$), for the smallest chain lengths,  $L < \bar \ell$, the distribution  shows a Gaussian profile. 
In this regime, the dynamics of the corresponding ABP model remains dominated by the translational diffusion,  
and $\la \rv^2 \ra \approx 2 d D t$. 
Equivalently, the polymer conformations remain dominated by the bond length fluctuations. 
With increasing contour length (time for ABP model), first the maximum at $\tilde r \approx 0$ starts to flatten as $L$ approaches $\bar \ell$~(Fig.\ref{fig:distb1}($b$)\,). 
For longer contours, $L = 10\, \bar\ell, \, 100\, \bar\ell$, the peak shifts towards
$\tilde r 
\approx 1$~(Fig.\ref{fig:distb1}($c$), ($d$)\,). The bending rigidity starts to dominate the polymer conformations in this regime. 
In this model, for the persistence to start to dominate the polymer morphology, 
a relatively long chain is required. This behavior contrasts the current model from the WLC polymer, where the chain transforms from a rigid rod to flexible chain behavior monotonically, with increasing chain length. 
For longer chains, the position of the peak in $p(\tilde r)$ fails to catch up to $L$ as a result of effective polymer softening. This behavior is reminiscent of the WLC polymer.  
At $L = 10^3  \bar \ell \equiv \ell_p$, the peak shifts to a shorter relative separation $\tilde r \approx 900 \bar \ell /L \lesssim 1$~(Fig.\ref{fig:distb1}($e$)\,).  For longest chains, $L \gg \ell_p$,  the distribution gets back to an approximate Gaussian shape with the maximum 
shifting back to $\tilde r = \ell_p/ L \approx 0$~(Fig.\ref{fig:distb1}($f$)\,). 
This regime corresponds to $\la \rv^2 \ra = 2 d \, D_{\rm eff} t$ of the ABP model.

\begin{figure}[!t]
\begin{center}
\includegraphics[width=8cm]{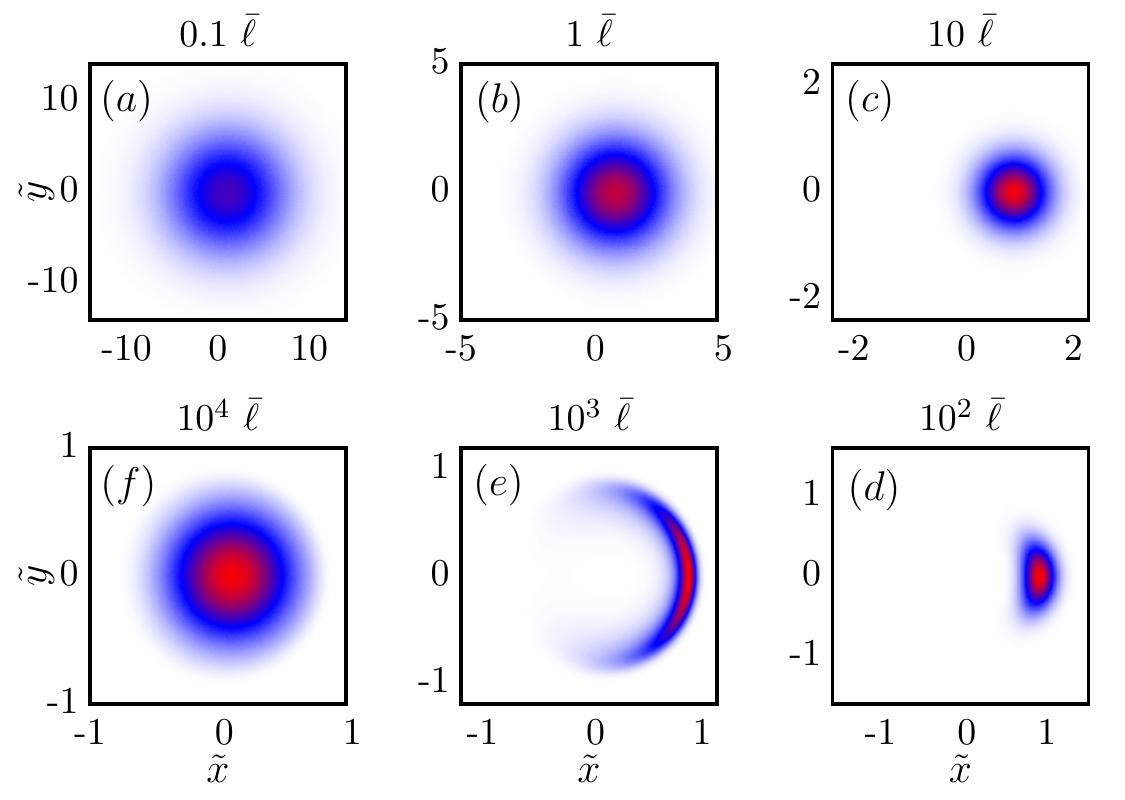}  
\caption{ (color online) The 2d displacement distributions $p(\tilde x, \tilde y)$ at $D_{r} = 1.0\,\t_u^{-1}$, $v_0=10.0\,\s/\t_u$ and $D=0.1\,\s^2/\t_u$ over different time segments $\t$,
presented as heat maps. The length of trajectories considered are $L = v_0 \t = 0.1 \,\bar \ell\,(a), 1.0\, \bar\ell\,(b), 10\,\bar\ell\,(c), \dots, 10^4 \bar \ell\,(f)$.
} 
\label{fig:distb2}
\end{center}
\end{figure}

\begin{figure}[!t]
\begin{center}
\includegraphics[width=8cm]{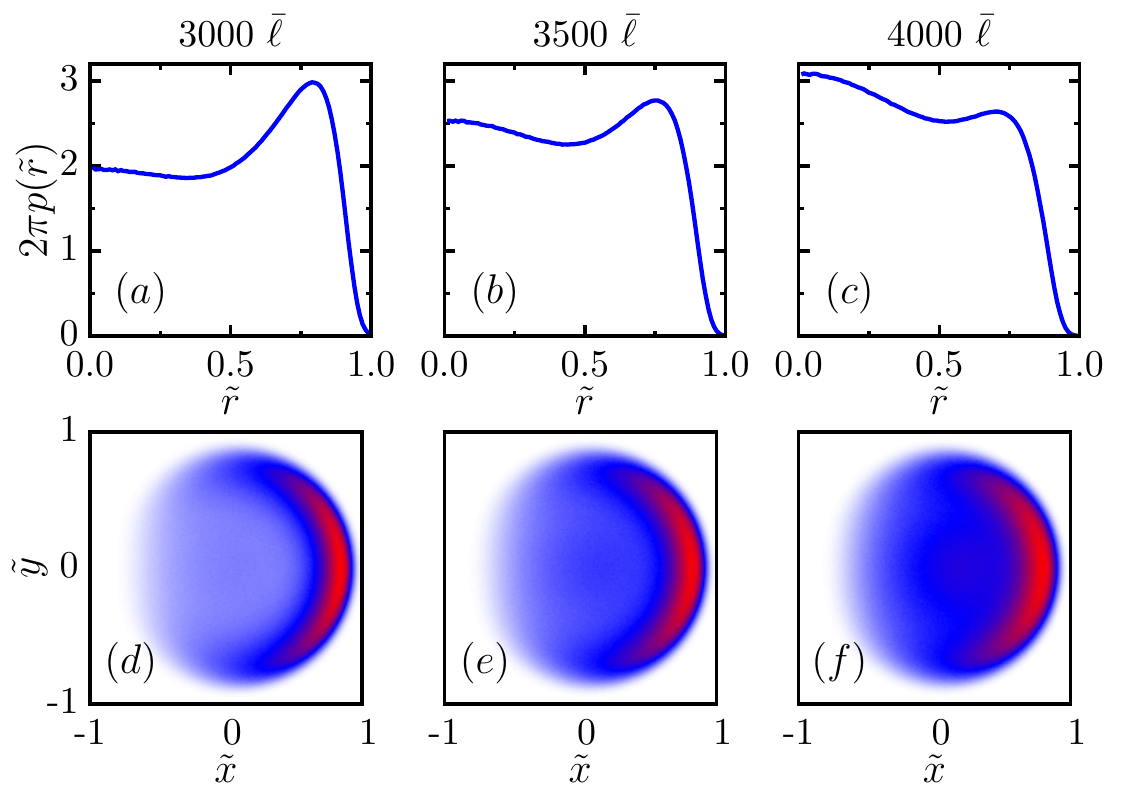}  
\caption{ (color online) The change in distributions $p(\tilde r)$ [$(a) - (c)$] and $p(\tilde x, \tilde y)$ [$(d) - (f)$] at $D_{r} = 1.0\,\t_u^{-1}$, $v_0=10.0\,\s/\t_u$ and $D=0.1\,\s^2/\t_u$ for different contour lengths $L=v_0 \t$ in the regime of bimodality. 
The persistence length $\ell_p = 10^3 \bar \ell$. The length of trajectories considered are $L=3 \ell_p\, [ (a), (d) ] \, 3.5 \ell_p\, [(b), (e)],\, 4 \ell_p\, [(c), (f)]$.
} 
\label{fig:distb3}
\end{center}
\end{figure}

Fig.\ref{fig:distb2} shows the full two-dimensional end-to-end distribution function $p(\tilde x, \tilde y)$ as a contour plot. Here $\tilde x = x/L$ and $\tilde y = y/L$. 
It is evident how the symmetry of the distribution changes with increasing 
contour-length of the polymer. With $L$, the peak shifts towards positive $x$-axis, the orientation of the first end of the polymer, but the distribution around the peak remains circularly symmetric  up to $L = 10\, \bar \ell$~(Fig.\ref{fig:distb2}($a$)--($c$)\,). 
Beyond this point, even around the peak, the distribution gets rotationally asymmetric, opening up as a partial ring-like structure at $L=\ell_p = 10^3 \bar \ell$~(Fig.\ref{fig:distb2}($d$)--($e$)\,). 
For the longest chain of $L=10^4 \bar \ell$, the distribution recovers its spherical symmetry and gets back to the Gaussian profile~(Fig.\ref{fig:distb2}($f$)\,).  
It is interesting to note that in terms of the rigidity parameter
$L/\ell_p$ the last two values of $L$ falls at $\ell_p$ and $10\,\ell_p$. For the chain in consideration, the effective spring-stiffness of the bonds $A=v_0/2D = 50 \s^{-1}$ is large enough to suppress bond-length fluctuations to within $7\%$, allowing an approximate  
WLC description of the effective chain. The WLC polymer is known to show a rigid rod to Gaussian transition mediated by a bistable region between $1 < L/\ell_p < 10$ (near $L/\ell_p \approx 3-4$)~\cite{Dhar2002, Chaudhuri2007}.   
 We find a similar transition through bistability at $L/\ell_p = 3.5$~(Fig.\ref{fig:distb3}). The distributions obtained in Fig.\ref{fig:distb3} are reminiscent of the property of  WLC polymer with one end tethered towards a fixed orientation~\cite{Chaudhuri2007}.

\section{Comparison of ABP with AOUP}
\label{sec:aoup}
In experiments often one encounters a question as to which model is better suited  to describe the observed spatiotemporal behavior of self propelled colloids~\cite{Kurzthaler2018}. Active Brownian particles (ABP), or a related model of active Ornstein-Uhlenbeck process (AOUP) are  used sometimes for such descriptions. The over-damped motion in AOUP model is given  by~\cite{Das2018}
\bea
\dot \rv &=& \vv + \sqrt{2D}\, \bm{\eta}(t) \nn\\
\dot \vv &=& -\g \vv + \sqrt{2 D_v}\, \bm{\eta}^v(t).
\label{eq:aoup}
\eea
The Gaussian random processes are defined by $\la \eta_i \ra = 0$, $\la \eta_i (t) \eta_j (t') \ra = \d_{ij} \d (t-t')$, 
$\la \eta^v_i \ra = 0$, $\la \eta^v_i (t) \eta^v_j (t') \ra = \d_{ij} \d (t-t')$, where $i,\,j$ denote components of the vectors. 
We assume ${\bm \eta}$ and ${\bm \eta^v}$ to be independent random processes. 
Given the Gaussian nature of the AOUP, it is straightforward to derive analytic expressions, including the probability distributions describing its dynamics~\cite{Das2018}. 
The distribution function $p(\rv (t),\vv(t), t\, | \,\rv (0),\vv(0)\,)$ for a given initial condition denoted by $\rv (0),\vv(0)$ can be obtained from the knowledge of the first two cumulants. 
Directly solving Eq.(\ref{eq:aoup}) one can obtain the moments  
\begin{align}
&\la \rv \ra = \f{ \vv (0) }{\g} \left( 1- e^{-\g t}\right), \label{eq:rv_aoup}\\
&\la \vv \ra =  \vv (0) \,  e^{-\g t}, \label{eq:vv_aoup}\\
&\la \rv^2 \ra =  2 d D t + \f{2 d D_v}{\g^3} \left[\g t - (1-e^{-\g t}) \right] \nn\\
 &~~~~~~~ +   \la \rv \ra^2- \f{d D_v}{\g^3} \left( 1- e^{-\g t} \right)^2, \label{eq:rv2_aoup} \\
&\la \vv^2 \ra =  \vv^2(0)  \, e^{-2 \g t} +  \f{d D_v}{\g}  \left( 1 - e^{-2 \g t} \right), \label{eq:vv2_aoup} \\
&\la \vv \cdot \rv \ra =  \f{ \vv^2(0) }{\g} \left(1 - e^{-\g t} \right) e^{-\g t}. 
\nn \\
&~~~~~~~~+ \frac{d D_v}{\gamma^2} (1-e^{-\gamma t})^2~. \label{eq:rvvv_aoup}
\end{align} 
We make the following identifications between AOUP and ABP parameters: 
\begin{align}
\vv^2(0)=\vv_0^2,~~~
\g = (d-1) D_r,~~~
\f{d D_v}{\g} =\vv_0^2. \label{eq:paraID}
\end{align}
Then we see that the  evolution of all the moments computed in Eqs.~(\ref{eq:rv_aoup},\ref{eq:vv_aoup},\ref{eq:rv2_aoup},\ref{eq:vv2_aoup},\ref{eq:rvvv_aoup})
have exactly the same form as those obtained for the ABP.
In particular we see that Eq.(\ref{eq:rv2_aoup})  simplifies to the form 
\bea
\la \rv^2 \ra =  2 d D t + \f{2 d D_v}{\g^3} \left[\g t - (1-e^{-\g t}) \right] .
\eea 
which can be compared with that for ABP obtained in Eq.(\ref{eq:r2}), when $\g$ and $D_v$ are interpreted using Eq.(\ref{eq:paraID}). 
 Similarly, after simplification $\la \vv^2 \ra = \vv^2(0)$, and $\la \vv \cdot \rv \ra = \left[ \vv^2(0) /\g \right] \left(1 - e^{-\g t} \right)$.

The ABP and AOUP models can thus be clearly distinguished only in terms of higher moments. 
Let us first evaluate  the fourth moment $\la \rv^4 \ra$ for a general Gaussian process (such as the AOUP) in terms of the lower order moments. For this, we write  $\rv = \d \rv + \la \rv \ra$ so that  
\begin{align}
& \la \rv^4 \ra = \la (\d r_i + \la r_i \ra)^2  \, (\d r_j + \la r_j \ra)^2 \ra \nn \\
& = \la \d r_i^2 \d r_j^2 \ra + 2 \la \d r_i^2 \ra \la r_j\ra^2 + 4 \la r_i \ra \la r_j \ra \la \d r_i \d r_j \ra + \la r_i \ra^2 \la r_j \ra^2 . \nn
\end{align}
Using Wick's theorem for Gaussian variables, 
\bea
\la \d r_i^2 \d r_j^2  \ra = \la \d r_i^2 \ra \la \d r_j^2 \ra + 2 \la \d r_i \d r_j \ra^2, \nn
\eea
we then get 
\bea
& \la \rv^4 \ra = \la \d \rv^2 \ra^2 + 2 \la  \d r_i \d r_j \ra^2 + 2 \la \d \rv^2 \ra \la \rv \ra^2 \nn\\
& +4 \la r_i \ra \la r_j \ra \la \d r_i \d r_j \ra + \la \rv \ra^4 .
\label{eq:r4aoup2}
\eea
This relation is true for any Gaussian process.  Let us define the functional on the right hand side of the above equation as a generalized moment for an arbitrary process, not necessarly Gaussian, and denote it by 
\begin{align}
\mu_4 &:=\la \d \rv^2 \ra^2 + 2 \la  \d r_i \d r_j \ra^2 + 2 \la \d \rv^2 \ra \la \rv \ra^2 \nn\\
& +4 \la r_i \ra \la r_j \ra \la \d r_i \d r_j \ra + \la \rv \ra^4 .
\label{eq:mu4def}
\end{align}
From our explicit solution for the ABP  and AOUP we find that 
\begin{align}
 \la \d r_i (t) \d r_j (t) \ra = 
\f{\d_{ij}}{d} \la \d \rv^2 \ra.  
\end{align}  
Replacing this relation in Eq.(\ref{eq:r4aoup2})
we obtain
{\color{violet}
\begin{align}
\mu_4 = \left( 1 + \f{2}{d}\right) \la \d \rv^2 \ra  \left( \la \d \rv^2 \ra + 2    \la  \rv \ra^2   \right)  +  \la \rv \ra^4.
\end{align}
}
Note that  for AOUP  we would have  $\la \rv^4 \ra = \mu_4$ but this would not be the case for ABP.

\begin{figure}[t]
\begin{center}
\includegraphics[width=8cm]{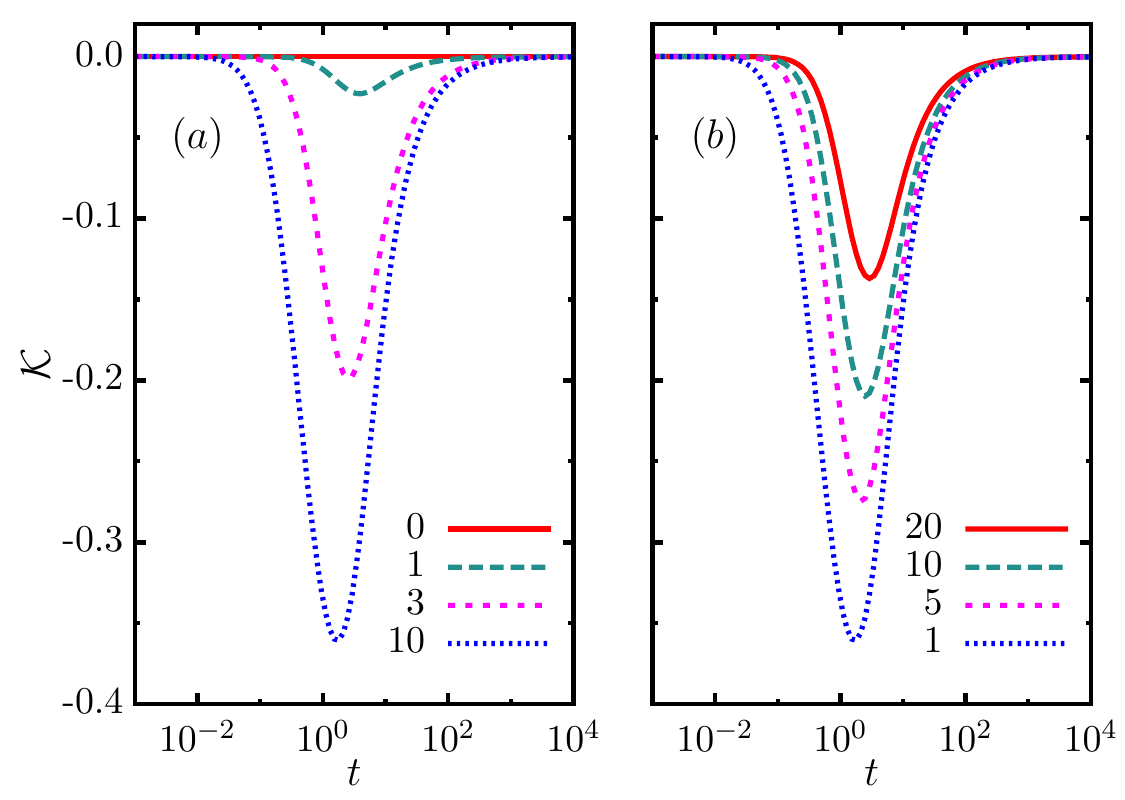} 
\caption{ (color online) Deviation from Gaussian nature in terms of ${\cal K}$ is shown as a function of time $t$ expressed in units of $\t_r = 1/D_r$. The translational diffusion $D = 1.0 \,\s^2/\t_u$ is held constant.  $(a)$\,Plot with $D_{r}=1.0\, \t_u^{-1}$ for $v_0 \, \t_u/\s=0$, $1$, $3$, $10$. $(b)$\,Plot at $v_0=10\,\s/\t_u$ for $D_{r}\t_u=1$, $5$, $10$, $20$.} 
\label{fig:kurtosis}
\end{center}
\end{figure}

In Eq.(\ref{eq:r4_abp}) we have already computed the explicit form of $\la \rv^4 \ra$  for ABP. It is then straightforward to evaluate the kurtosis in $d$-dimensions defined as 
\bea
{\cal K} = \f{\la \rv^4 \ra}{\mu_4} - 1 .
\eea
By definition, this quantity is identically zero for the AOUP.
In Fig.\ref{fig:kurtosis} we show plots of ${\cal K}$ obtained for the ABP model, using our analytical expressions
 for $\la \rv^4 \ra$, and that of $\la \d \rv^2 \ra$, and $\la \rv \ra$. 
The kurtosis was calculated numerically in earlier studies of ABP~\cite{sevilla2014,Das2018}. 
The two plots in Fig.\ref{fig:kurtosis} show variation of ${\cal K}$ with time for different amount of activity, measured in terms of active speed $v_0$~($a$) and rotational diffusion of 
active orientation $D_r$~($b$).  The plots in Fig.\ref{fig:kurtosis} show the time dependence 
of ${\cal K}$ at fixed translational diffusion $D = 1.0\,\s^2/\t_u$.  
At $v_0 \t_u/\s=0$, the ABP motion is the same as equilibrium diffusion showing ${\cal K}=0$ in Fig.\ref{fig:kurtosis}($a$).
With increasing $v_0$ the deviation from Gaussian nature characterized by the amplitude of  ${\cal K}$
becomes more pronounced and prevails for longer duration in time. 
Beyond $v_0 = 100\, \s/\t_u$ the Kurtosis touches a maximum amplitude of ${\cal K} \approx -0.4$, and the curve does not change appreciably with further increases in $v_0$.
On the other hand, as is shown in Fig.\ref{fig:kurtosis}($b$), the deviation from zero of ${\cal K}$ reduces with increasing orientational diffusion $D_r$, 
better randomizing the orientation of activity bringing the evolution back towards equilibrium behavior. 
Over long enough time the trajectories behave as that of diffusion, leading to ${\cal K} = 0$ for all $v_0$. 

The measure ${\cal K}$, in terms of $\la \rv^4 \ra$, $\la \d \rv^2 \ra$ and $\la \rv \ra$, is easily obtainable from observed trajectories of  self propelled colloids. 
This would suffice to deduce if the properties shown by a self propelled particle is better described by the AOUP or the ABP model, 
depending on whether ${\cal K}$ remains vanishingly small all through the evolution, or deviates from zero significantly in the intermediate time window as in ABP. 
Clearly, this measure requires much less information with respect to the measurement of the complete distribution functions proposed in Ref.\cite{Kurzthaler2018}.

\section{Discussion}
\label{sec:outlook}

In this paper, we studied free ABPs in the presence of translational thermal noise. 
We have established a mapping of the ABP trajectories 
to an equivalent polymer model. 
The bond stiffness and bending rigidity of the mapped polymer are determined by the active speed, orientational diffusion, and thermal noise in the ABP model.
In the limit of vanishing orientational diffusivity, the ABP trajectories map to a Gaussian polymer under directed external force.
The other limit of vanishing translational diffusion in the ABP, reduces the mapped polymer to the well known WLC model of the semiflexible 
chain. 
Comparisons of the distribution functions for non-equilibrium displacements in ABP, and the end-to-end separations in the 
equilibrium polymer model showed good agreement. 
Remarkably, with increasing trajectory length the mapped polymer undergoes {\em re-entrant} transitions from a Gaussian chain, to rigid filament, back to a Gaussian chain via
a pronounced bimodality which is a characteristic of the semiflexible polymer.  

Secondly, we have shown how arbitrary moments of the position and active orientation vectors of ABPs in arbitrary dimensions can be calculated using the governing Fokker-Planck equation.  
For this we  utilized a Laplace transform approach used earlier for the WLC model~\cite{Hermans1952}.
Our calculation differs significantly from other recent analytic approaches employed for ABPs~\cite{sevilla2015,Basu2018, Malakar2018}.
The expressions for moments that we obtained were compared against numerical simulations, and have been utilized to analyze all the observed dynamical crossovers. 
Finally, we derived an analytical expression for the kurtosis of position vector for ABPs, and have shown how it differs from a related AOUP model. 
This can be utilized to analyze observed trajectories of self propelled colloids, to identify if they can be described by the AOUP, or are better
described by the ABP model.


\acknowledgements
The computations were supported in part by SAMKHYA, the high performance computing facility at Institute of Physics,
Bhubaneswar.
D.C. thanks SERB, India for financial support through grant number MTR/2019/000750.  
A.D. acknowledges support of the Department of Atomic Energy, Government of India, under project no.12-R$\&$D-TFR-5.10-1100.  This research was supported in part by the International Centre for Theoretical Sciences (ICTS) during a visit for participating in the program - Thirsting for Theoretical Biology (Code: ICTS/ttb2019/06).


\bibliographystyle{prsty}

\end{document}